\documentclass[5p,english,authoryear]{elsarticle}
\usepackage{url}
\usepackage{amsmath}
\usepackage[english]{babel}

\journal{Astronomy \& Computing}

\newcommand{\figureref}[1]{Figure \ref{#1}}
\newcommand{\tableref}[1]{Table \ref{#1}}
\newcommand{\code}[1]{{\normalfont\fontfamily{cmvtt}\selectfont #1}}

\begin{document}

\begin{frontmatter}

\title{The moving mesh code \textsc{Shadowfax}}
\author[adres]{B.~Vandenbroucke\corref{cor1}}
\ead{bert.vandenbroucke@ugent.be}
\author[adres]{S.~De~Rijcke}
\ead{sven.derijcke@ugent.be}

\cortext[cor1]{Corresponding author}
\address[adres]{Dept. Physics \& Astronomy, Ghent University, Krijgslaan 281,
S9, 9000 Gent, Belgium}

\begin{abstract}
We introduce the moving mesh code \textsc{Shadowfax}, which can be used to
evolve a mixture of gas, subject to the laws of hydrodynamics and gravity, and
any collisionless fluid only subject to gravity, such as cold dark matter or
stars. The code is written in C++ and its source code is made available to the
scientific community under the GNU Affero General Public License. We outline the
algorithm and the design of our implementation, and demonstrate its validity
through the results of a set of basic test problems, which are also part of the
public version. We also compare \textsc{Shadowfax} with a number of other
publicly available codes using different hydrodynamical integration schemes,
illustrating the advantages and disadvantages of the moving mesh technique.
\end{abstract}


\end{frontmatter}

\section{Introduction}

Modern simulations of galaxy formation and evolution crucially depend on an
accurate treatment of the hydrodynamics of the interstellar medium (ISM)
\citep{2014Vogelsberger, 2015Schaye}. The ISM fuels star formation and is
disrupted by stellar feedback, and it is this complex interplay that at least
partly governs the observable content of galaxies \citep*{2015Verbeke}. If we
want to be able to compare simulated galaxies with observations, we need to
properly resolve these effects.

Hydrodynamics is also important on smaller scales, when simulating star-forming
clouds \citep{2011Greif, 2015Dobbs}, feedback from a single star
\citep{2015Geen}, or even planet formation in a circumstellar disc
\citep{2012Duffell}. A robust hydrodynamical integration scheme, optionally
extended with magnetic fields, self-gravity or radiation transport, is hence an
indispensable tool for many astrophysical simulators.

Historically, two major classes of hydrodynamical solvers have been developed~:
grid based Eulerian techniques \citep{2002Teyssier, 2012Keppens}, and
particle-based Lagrangian techniques \citep{2005Springel, 2012Price}. Both
discretize the fluid as a finite set of fluid elements. In the former, the fluid
elements are cells, usually defined through a (hierarchical) Cartesian grid,
which have a fixed position in space, but can be allowed to refine or derefine
according to the quality of the integration. In the latter, the fluid elements
are particles, which move along with the flow, with the hydrodynamics being
expressed as inter-particle forces. It is generally acknowledged that grid based
Eulerian techniques are more accurate at solving the equations of hydrodynamics,
especially since many particle-based implementations have fundamental
difficulties in resolving hydrodynamical instabilities \citep{2007Agertz}.
Nonetheless, Lagrangian techniques are widely used to simulate systems with a
high dynamic range, like cosmological simulations and simulations of galaxies,
since they more naturally concentrate computational resources on regions of
interest, and provide a Galilean invariant reference frame.

Recently, a new class of hydrodynamical solvers has been developed, mainly
through the work of \citet{2010Springel}, which aims to combine the advantages
of Eulerian and Lagrangian techniques \citep*[see also][]{2011Duffell,
2015Yalinewich}. This new technique uses a moving grid to discretize the fluid,
and combines an unstructured grid based finite volume integration scheme with
the Lagrangian nature of a particle method. We will refer to this method as a
\emph{moving mesh technique}.

A number of moving mesh codes are presented in the literature
\citep{2010Springel, 2011Duffell}, but only two of them are
publicly available~: \textsc{rich}\footnote{ascl:1410.005}
\citep*{2015Yalinewich}, written in C++, and
\textsc{FVMHD3D}\footnote{\url{https://github.com/egaburov/fvmhd3d}}, written in
the parallel object-oriented language \textsc{Charm++} \citep*{2012Gaburov}.

\begin{figure}
\centering{}
\includegraphics[width=0.4\textwidth]{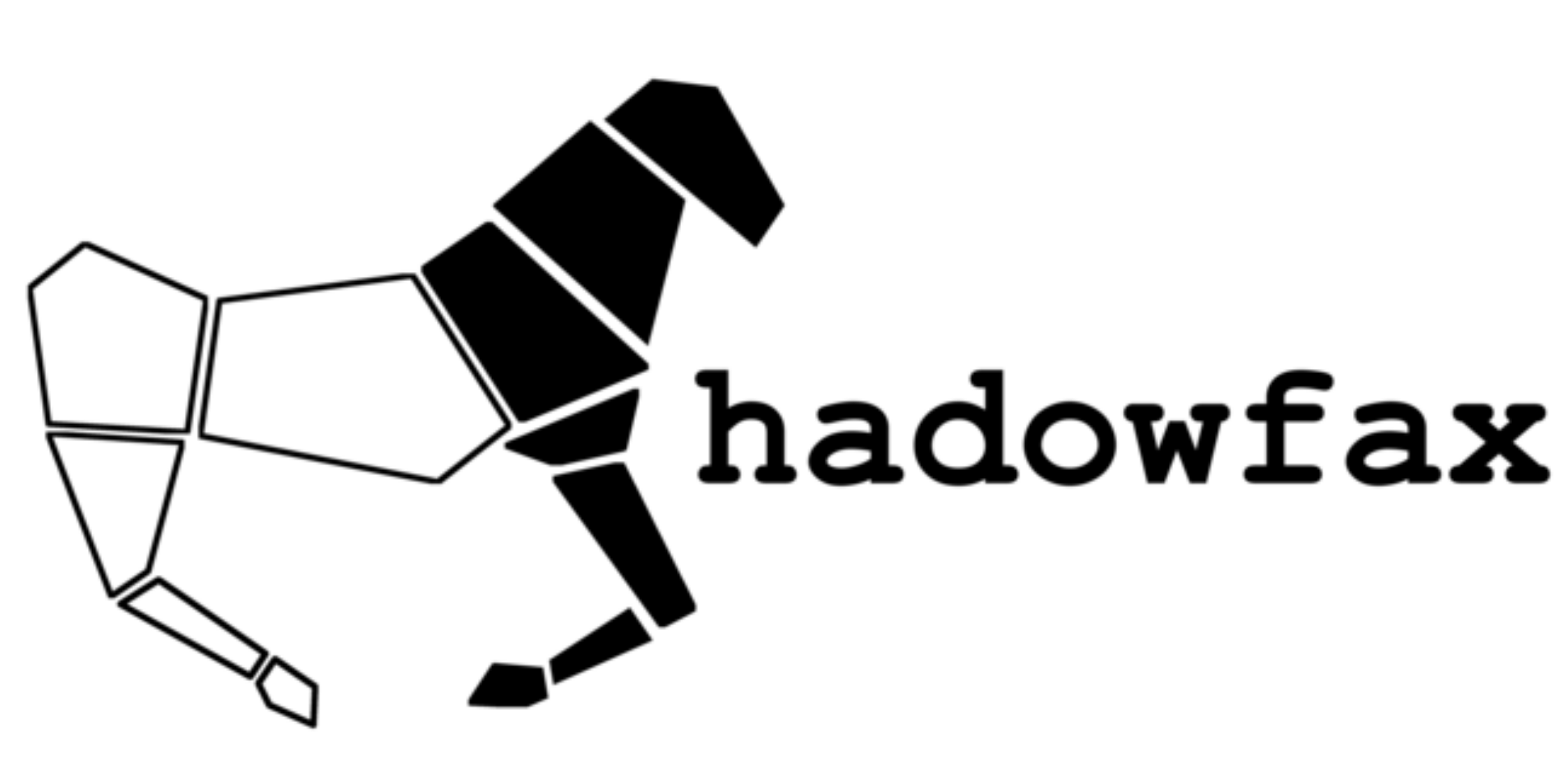}
\caption{The \textsc{Shadowfax} logo.\label{fig_logo}}
\end{figure}

In this paper, we introduce the new, publicly available moving mesh code
\textsc{Shadowfax} (the logo of the code is shown in \figureref{fig_logo}).
\textsc{Shadowfax} is written in C++, and makes ample use of the object-oriented
capabilities of the language to provide an easy to extend framework. The code is
parallelized for use on distributed memory systems using the Message Passing
Interface (MPI)\footnote{\url{http://www.mpi-forum.org}}, and makes use of the
open source \textsc{Boost} C++ libraries\footnote{\url{http://www.boost.org}} to
extend basic C++ language features. The code supports input and output using the
HDF5 library\footnote{\url{https://www.hdfgroup.org/HDF5}} in a format
compatible with the output of \textsc{Gadget2}\footnote{ascl:0003.001}
\citep{2005Springel}, \textsc{gizmo}\footnote{ascl:1410.003} \citep{2015Hopkins}
and \textsc{swift}\footnote{\url{http://icc.dur.ac.uk/swift/}}
\citep{2013Gonnet}. A user friendly compilation process is guaranteed through
the use of \textsc{CMake}\footnote{\url{https://cmake.org}}.

The hydrodynamical algorithm implemented in \textsc{Shadowfax} is the same as
described by \citet{2010Springel}, but with an additional per-face slope limiter
and flux limiter, and optional alternative approximate Riemann solvers. The
gravitational calculation is the same as the tree force calculation in
\textsc{Gadget2} \citep{2005Springel}, and uses the same relative tree opening
criterion and Ewald summation technique for periodic boundary conditions. We
have ported this algorithm to an object-oriented version, which makes use of
compile-time polymorphism using C++ templates. This ensures a clear separation
of the algorithmic details underlying the tree walk from the actual physics
involved with the gravitational calculation. This way, it is much easier to
focus on one particular aspect of the code, e.g. scalibility, precision...,
without needing to worry about other aspects.

Likewise, we have separated the geometrical details contained in the moving mesh
from the hydrodynamical integration as mush as possible, to make it easier to
replace parts of the algorithm (e.g. the Riemann solver, the grid...) by simply
implementing an alternative class.

Our code is predominantly meant to be used in astrophysical simulations of
galaxy formation and evolution, but could have applications in other areas of
science as well, as it is not difficult to replace the Euler equations of
hydrodynamics by e.g. the shallow water equations by implementing a different
Riemann solver. Furthermore, the Voronoi grid used to discretize the fluid can
also be used for other purposes, e.g. for the suppression of Poisson noise in
randomly sampled distributions through Lloyd's algorithm \citep{1982Lloyd}, or
as density estimator in N-body simulations \citep{2014CloetOsselaer}.

In this paper, we outline the basic working of \textsc{Shadowfax}. We mainly
focus on the C++ implementation and the object-oriented design of our code, and
compare our code with other hydrodynamical solvers on a number of test problems.
Although the current version of \textsc{Shadowfax} focusses more on design and
accuracy than on performance, we also highlight some basic strong and weak
scaling tests. Performance optimizations and extra physical ingredients (e.g.
gas cooling, star formation and stellar feedback...) will be added in future
versions of the code. The source code of \textsc{Shadowfax} is publicly
available from \url{https://github.com/AstroUGent/shadowfax}, and is distributed
under the GNU Affero General Public
License\footnote{\url{http://www.gnu.org/licenses}}.

\section{Algorithm}

Many of the algorithms implemented in \textsc{Shadowfax} were already discussed
in \citet{2005Springel} and \citet{2010Springel}. For completeness, we summarize
them below and point out the differences where necessary.

\textsc{Shadowfax} is based on a finite volume method, which subdivides the
computational box into a (large) number of small cells. The hydrodynamical
integration is governed by the exchange of fluxes between these cells.

These fluxes involve the \emph{conserved variables}~: mass ($m$), momentum
($\boldsymbol{p}$) and total energy ($E$). The Euler equations of hydrodynamics
however are usually formulated in terms of \emph{primitive variables}: density
($\rho{}$), flow velocity ($\boldsymbol{v}$) and pressure ($p$). The pressure is
sometimes replaced by the thermal energy ($u$) or some form of entropic function
of the fluid, by using the \emph{equation of state} of the fluid. In this work,
we will always assume an ideal gas, with an equation of state of the form
\begin{equation}
p = (\gamma{}-1)\rho{}u,
\end{equation}
where $\gamma{}$ is the \emph{adiabatic index} of the gas, for which we will
adopt the value $\gamma{}=5/3$, unless otherwise stated.
The conserved variables and primitive variables can be converted into one
another whenever a volume ($V$) is available, since
\begin{align}
m &= \rho{}V \\
\boldsymbol{p} &= m\boldsymbol{v} \\
E &= m u + \frac{1}{2}m\boldsymbol{v}^2.
\end{align}

It is common practice to combine the conserved and primitive variables into two
\emph{state vectors},
\begin{equation}
\boldsymbol{Q} = \begin{pmatrix} m \\ \boldsymbol{p} \\ E \\ \end{pmatrix}
\quad{}\textrm{and}\quad{}
\boldsymbol{W} = \begin{pmatrix} \rho{} \\ \boldsymbol{v} \\ p \end{pmatrix}.
\end{equation}
The change in conserved variables $\boldsymbol{Q}_i$ for a cell $i$, during an
integration time step of length $\Delta{}t$, is then given by
\begin{multline}
\Delta{}\boldsymbol{Q}_i = -\Delta{}t \sum_j A_{ij} \boldsymbol{F}_{ij}
\left(\boldsymbol{W}_i, \boldsymbol{W}_j, \nabla{}\boldsymbol{W}_i, \right.\\
\nabla{}\boldsymbol{W}_j, \left. \boldsymbol{v}_{ij}, \Delta{}t\right),
\end{multline}
where $A_{ij}$ is the surface area of the interface between cell $i$ and cell
$j$. $\boldsymbol{F}_{ij}$ is the flux between cell $i$ and cell $j$, which in
general depends on the primitive variables of both cells and their gradients,
the velocity $\boldsymbol{v}_{ij}$ of the face with respect to a frame of
reference fixed to the simulation box, and the integration time step.

When formulated in this way, the finite volume method can be applied to any
discretization of the fluid, as long as this discretization yields volumes to
convert conserved variables to primitive variables, and defines a concept of
neighbour relations between cells, and an associated surface area and velocity
for the neighbour interface. It can even be applied to mesh-free, particle-based
methods \citep{2015Hopkins}.

\begin{figure*}
\centering{}
\includegraphics[width=\textwidth]{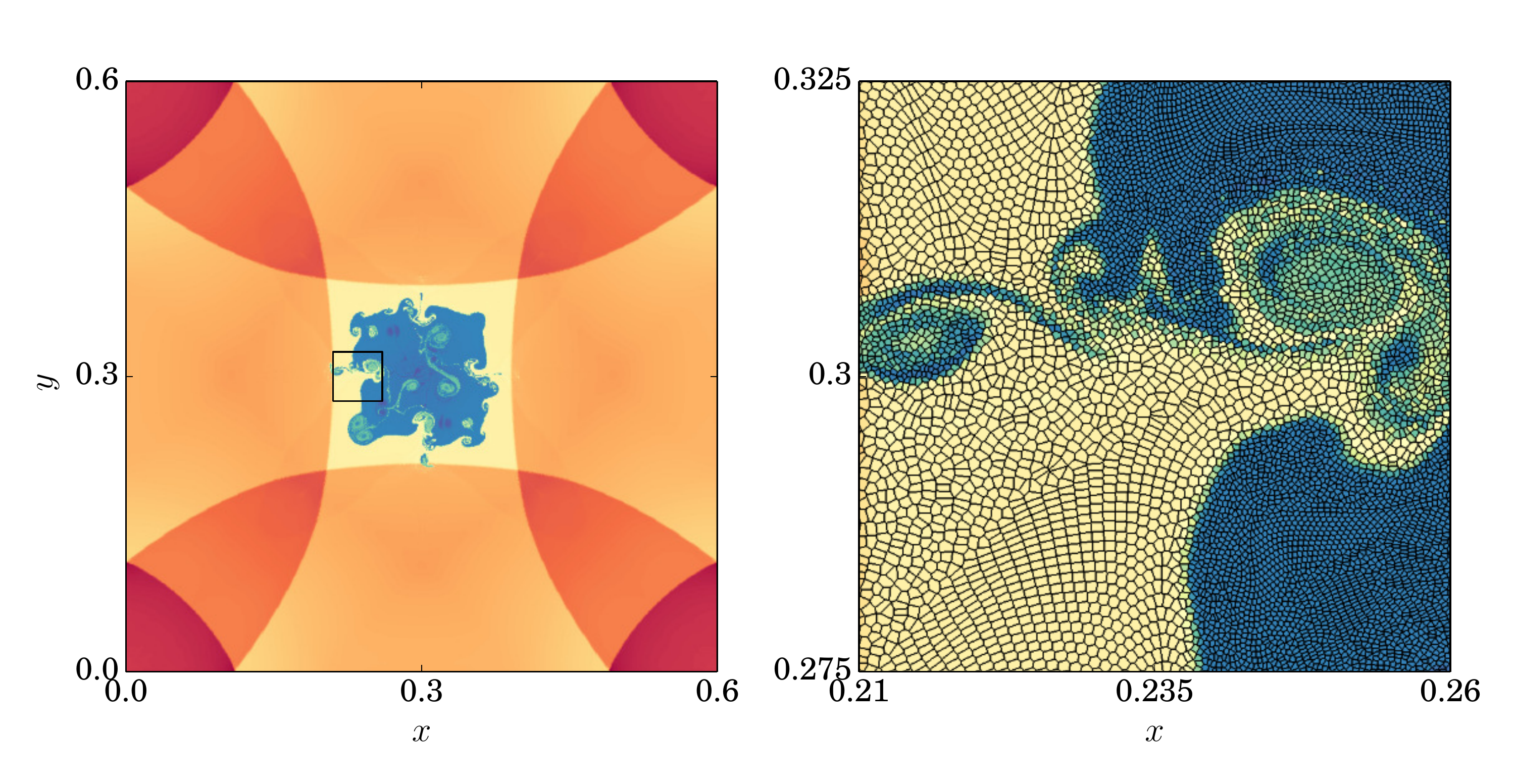}
\caption{Density colour plot for the 2D \citet{2003Liska} implosion test at
$t=0.5$. \emph{Left:} general view of the entire simulation box, \emph{right:}
zoom in on part of the central low density region. The Voronoi mesh used for the
discretization of the gas has been overplotted.\label{fig_implosion_zoom}}
\end{figure*}

In the case of a moving mesh method, the discretization is given by an
unstructured Voronoi mesh, a 2D example of which is shown in
\figureref{fig_implosion_zoom}. The mesh is defined by means of a set of mesh
generating points (\emph{generators}), with the cell associated with a specific
generator containing the region of space closest to that generator. A Voronoi
mesh can be defined in $D$ dimensions, but we will focus on the cases $D=2$ and
$D=3$. The Voronoi mesh has the interesting property that it is relatively
stable under small movements of the generators \citep{2011Reem}, so that cells
deform continuously under a continuous movement of the generators. We exploit
this property to allow the mesh to move in between integration time steps. The
surface area of the faces will change linearly in between steps, so that the
time averaged flux over the entire time step will be correct, even if it is
calculated at one specific moment in time.

By setting the velocities of the mesh generators equal to the local flow
velocity, the mesh will effectively move along with the fluid, and we end up
with a Lagrangian method. We can even reinterpret the generators as being
particles, so that the moving mesh technique becomes a true alternative for
particle-based methods. However, the underlying integration scheme uses the full
strength of a finite volume method, and hence will be more accurate. Note that
the quality of the integration will depend on the shape of the cells, with
highly irregular cells leading to less accuracy \citep{2012Vogelsberger}. To
ensure cell regularity, it is sometimes necessary to add extra correction terms
to the generator velocities. We employ the scheme of \citet{2010Springel} and
steer the cell generator towards the centroid of its cell if the distance
between generator position and cell centroid exceeds a fraction of the generic
cell size, i.e. the radius of a sphere with the same volume as the cell.

In the remainder of this section, we describe the mesh construction and flux
calculation in more detail, to introduce the concepts that are used in the
discussion of the \textsc{Shadowfax} implementation.

\subsection{Mesh construction}

We construct the Voronoi mesh through its dual Delaunay triangulation. The
latter is constructed using an incremental construction algorithm. In the
default implementation, the relevant parts of the mesh are reconstructed for
every time step. We also experimented with a mesh evolution algorithm
(Vandenbroucke \& De Rijcke, in preparation), which evolves the mesh instead of
reconstructing it.

As \citet{1997Shewchuk} pointed out, incremental Delaunay construction
algorithms can become unstable due to numerical round-off error. To prevent this
from happening, we employ arbitrary exact arithmetics for all geometrical tests
involved. Since the predicates of \citet{1997Shewchuk} depend on a number of
assumptions on the internal CPU precision that are not met on all hardware
architectures, we use the technique outlined by \citet{2010Springel}~: we map
the floating point coordinates of the mesh generators to the interval $[1,2]$
and use an integer representation of the mantissa to exactly calculate the
result of a geometrical test if the numerical error could lead to a wrong
result. We pre-calculated the maximal size of an integer necessary to store the
exact result and use the \textsc{Boost Multiprecision}
library\footnote{\url{
http://www.boost.org/doc/libs/release/libs/multiprecision/}} to perform the
calculations using long integer arithmetics.

The result of the mesh construction is a list of neighbours for every mesh
generator, and an associated list of faces, with each face consisting of an
ordered list of vertex positions. We need to compute the volume and geometrical
centroid of each cell from this, as well as the surface area and midpoint of
every face.

In 2D, the faces each contain only two vertices, with the midpoints being the
midpoints of the line segments formed by these two vertices. In this case, we
order the neighbours and faces counter-clockwise around the position of the
generator of the cell. The volume (2D surface area) of the cell is then given by
the sum of the surface areas of the triangles that are formed by the first
vertex and two consecutive other vertices. The centroid of the cell is then the
weighted average of the centroids of these same triangles.

In 3D, we use a similar technique to obtain the surface area and midpoint of the
individual faces. The total volume of the cell is then the sum of the volumes of
the pyramids with the faces as base and the cell generator as top (the latter
being guaranteed to lie inside the cell). The centroid is the weighted average
of the centroids of these pyramids.

The mesh construction algorithm only works if all cells, including those at the
borders of the simulation volume, have boundaries. To make sure this is the
case, we always enclose all generators in a simulation box, which is a cuboid.
Cells at the boundaries are then completed by inserting ghost generators. If we
want periodic boundaries, we insert periodic copies of generators at the other
side of the simulation box. We can also mirror the generator positions
themselves with respect to the boundary face of the simulation box, in this case
we have reflective boundaries.

During the flux calculation, these ghost generators then either represent the
cell of which they are a periodic copy, or the border cell itself, but with the
sign of the flow velocity along the boundary face normal reversed.

\subsection{Flux calculation}

We employ a Monotonic Upwind Scheme for Conservation Laws (MUSCL) combined with
a Hancock prediction step to estimate the fluxes. This scheme is more commonly
referred to as a MUSCL-Hancock scheme. The flux calculation consists of a cell
based interpolation and integration step, whereby the primitive variables at the
center of the cell (the centroid of the Voronoi cell) are interpolated to the
position of the midpoint of the face using the gradients of the primitive
variables in that cell, and are predicted forward in time for half the time step
using the same gradients and the Euler equations in primitive form.

These face reconstructed primitive variables are then used as the input for a
Riemann solver, to obtain appropriately averaged primitive quantities at the
interface which take into account the local wave structure of the Euler
equations \citep{2009Toro}. This can be done using either an exact, iterative
Riemann solver, or an approximate Riemann solver. The solution of the Riemann
problem, $\boldsymbol{W}_*$, is then used to calculate the hydrodynamical fluxes
as
\begin{equation}
\boldsymbol{F}(\boldsymbol{W}_*) = \begin{pmatrix}
\rho{}_* \boldsymbol{v}_* \\
\rho{}_* \boldsymbol{v}_* \boldsymbol{v}_* + p_* \\
\rho{}_* (u_* + \frac{1}{2}\boldsymbol{v}_*^2)\boldsymbol{v}_* +
p_*\boldsymbol{v}_*
\end{pmatrix}.
\end{equation}

The Riemann problem is generally formulated in 1D, with the change in variables
corresponding to a change in $x$ coordinate. In an unstructured mesh, the face
normals are generally not aligned with the $x$ axis, so that we need to rotate
the primitive variables to a reference frame where the $x$ axis is aligned with
the surface normal of the face. This only affects the fluid velocity components.
The solution of the Riemann problem then needs to be rotated back to the
original reference frame.

When the cells are allowed to move, the faces will move as well, and we need to
transform to a frame of reference moving along with the face before solving the
Riemann problem. This again only affects the fluid velocity components of the
primitive variables.

However, the flux needs to be adapted in this case as well~: even if the
hydrodynamical flux through the face would normally be zero, there will be net
flux, caused only by the movement of the cell. This flux is given by
\begin{equation}
\boldsymbol{F}_\textrm{mov}(\boldsymbol{W}_*) = -\begin{pmatrix}
\rho{}_* \boldsymbol{v}_{ij} \\
\rho{}_* \boldsymbol{v}_* \boldsymbol{v}_{ij} \\
\rho{}_* (u_* + \frac{1}{2}\boldsymbol{v}_*^2)\boldsymbol{v}_{ij}
\end{pmatrix}.
\end{equation}
We can either add this correction flux to the hydrodynamical flux
\citep{2010Springel}, or calculate the hydrodynamical flux in a reference frame
moving along with the face and correct for the movement afterwards
\citep*{2011Pakmor}.

\subsection{Limiters}

A side effect of the gradient interpolation step discussed above is the possible
introduction of new extrema of the primitive variables, which can cause spurious
oscillations in the solution \citep{2009Toro}. These spurious oscillations can
be avoided by constructing monotone schemes, whereby the gradients of a cell are
limited so that no interpolated primitive value can exceed the primitive
variables in one of the neighbouring cells. However, such monotone schemes are
no longer second order accurate in space, so that a limiting procedure
inevitably leads to a loss of accuracy. It might therefore be better to use a
somewhat less restrictive gradient limiter, as long as the spurious oscillations
do not dominate the local solution \citep{2015Hopkins}. We use the cell wide
slope limiter described by \citet{2010Springel}.

Apart from a cell wide slope limiting procedure that limits the gradients of the
cell during the gradient calculation (which is performed after the primitive
variables have been calculated, but before the fluxes are computed), it is also
possible to limit the interpolated values at the faces. The general idea of such
a pair-wise slope limiter is to conserve the wave structure of the Riemann
problem. This wave structure generally consists of a central contact
discontinuity and a left and right wave, which can be either a shock wave or a
rarefaction wave \citep{2009Toro}. Spurious oscillations arise when the solution
of the Riemann problem with the cell centered primitive variables as input
yields e.g. a left and right rarefaction wave, while the Riemann problem with
the interpolated variables at the face yields a left shock wave and a right
rarefaction wave. The left shock is not present in the first order solution, and
introduces a growing artefact in the solution. With a pair-wise limiter, we
limit the left and right interpolated values in such a way that the original
wave structure of the Riemann problem is the same, but the input values can
still differ from the cell centred values. We implemented the pair-wise slope
limiter of \citet{2015Hopkins}.

Even with appropriate slope limiters, it is still possible that the calculated
fluxes are too large, i.e.\ exceed the value of the conserved variables in the
cell. For the mass and energy of the cell, this is fatal, since this can cause
negative masses and energies, which are evidently unphysical. This can happen
for example if the integration time step is too large, if the gradient
interpolation is done in an asymmetric way (due to pair-wise limiting), or if
external forces (e.g. gravity) contribute to the flux. The former is normally
excluded by choosing an appropriate time step criterion. To prevent the latter
from crashing the code, we implemented a \emph{flux limiter}, which ensures that
the flux through a face can never be larger than a fraction of the value of the
conserved variables inside the cell. This fraction is equal to the ratio of the
surface area of the face to the total surface area of the entire cell.

We note that a flux limiter is only used to ensure code stability, and in this
sense is equal to resetting the mass or energy of a cell to some very small
value whenever they become negative. However, by limiting the flux and not the
cell quantities themselves, we ensure manifest conservation of mass and energy,
which would otherwise be violated.

\subsection{Gravity}

Gravity is added as an extra term in the momentum equation:
\begin{multline}
\Delta{}\boldsymbol{p}_i = -\Delta{}t \sum_j A_{ij} \boldsymbol{F}_{ij,
\boldsymbol{p}} \\- \frac{1}{2}\Delta{}t \left( m_{i, \text{old}}
\boldsymbol{\nabla{}}_i \Phi{}_\text{old} + m_{i, \text{new}}
\boldsymbol{\nabla{}}_i \Phi{}_\text{new} \right),
\end{multline}
where $m_{i,\text{old}}$ and $m_{i,\text{new}}$ represent the mass inside the
cell before and after the update of the mass respectively, and
$\boldsymbol{\nabla{}}_i \Phi{}_\text{old}$ and $\boldsymbol{\nabla{}}_i
\Phi{}_\text{new}$ represent the gravitational acceleration before and after the
generator positions have been updated. $\boldsymbol{F}_{ij, \boldsymbol{p}}$ is
the hydrodynamical flux for the momentum, as given above.

The gravitational acceleration is also taken into account during the half step
prediction, before the flux calculation, and only affects the velocity.

Gravity also affects the total energy of the cell. Simply adding a term to the
energy equation does not take into account the movement of the mass that fluxes
through cell faces during the time step, and leads to significant energy errors.
We therefore use the following more involved equation to update the energy of a
cell:
\begin{multline}
\Delta{}E_i = -\Delta{}t \sum_j A_{ij} \boldsymbol{F}_{ij, E} \\ - \frac{1}{2}
\left( m_{i, \text{old}} \boldsymbol{w}_{i, \text{old}} \boldsymbol{\nabla{}}_i
\Phi{}_\text{old} + m_{i, \text{new}} \boldsymbol{w}_{i, \text{new}}
\boldsymbol{\nabla{}}_i \Phi{}_\text{new} \right) \\ -\frac{1}{4} \sum_j
\Delta{}m_{ij} \left( \boldsymbol{r}_i - \boldsymbol{r}_j \right)
\left( \boldsymbol{\nabla{}}_i \Phi{}_\text{old} + \boldsymbol{\nabla{}}_i
\Phi{}_\text{new} \right),
\end{multline}
where $\boldsymbol{w}_{i,\text{old}}$ and $\boldsymbol{w}_{i,\text{new}}$
represent the generator velocities before and after the update of the generator
positions, and $\Delta{}m_{ij}$ is the mass that fluxed from cell $i$ to cell
$j$ during the time step. The sum extends over all neighbours of the cell.

We soften the gravitational acceleration for both the hydrodynamical and
collisionless component using a spline kernel with fixed softening length
\citep{2005Springel}. A gravitational time step criterion based on the size of
the gravitational acceleration and the softening length is combined with the
hydrodynamical time step criterion to set the particle time steps.

\section{Implementation}

The finite volume method and the actual discretization of the fluid as a Voronoi
mesh are clearly well separated concepts, so that it makes sense to separate
them in the design of a moving mesh code. It then becomes possible to test the
finite volume method using e.g. a fixed Cartesian mesh, which is computationally
much cheaper to construct, or to treat the mesh generators as particles and
calculate volumes and interfaces using a mesh-free method \citep{2015Hopkins}.

Furthermore, various parts of the finite volume method can be adapted, and could
be considered to be run time parameters for the code~: the choice of Riemann
solver, the choice of slope limiter,... A good code design should make it
possible to easily exchange these components without affecting other parts of
the algorithm, and where possible also without the need to recompile the code.

\begin{figure}
\centering{}
\includegraphics[width=0.5\textwidth]{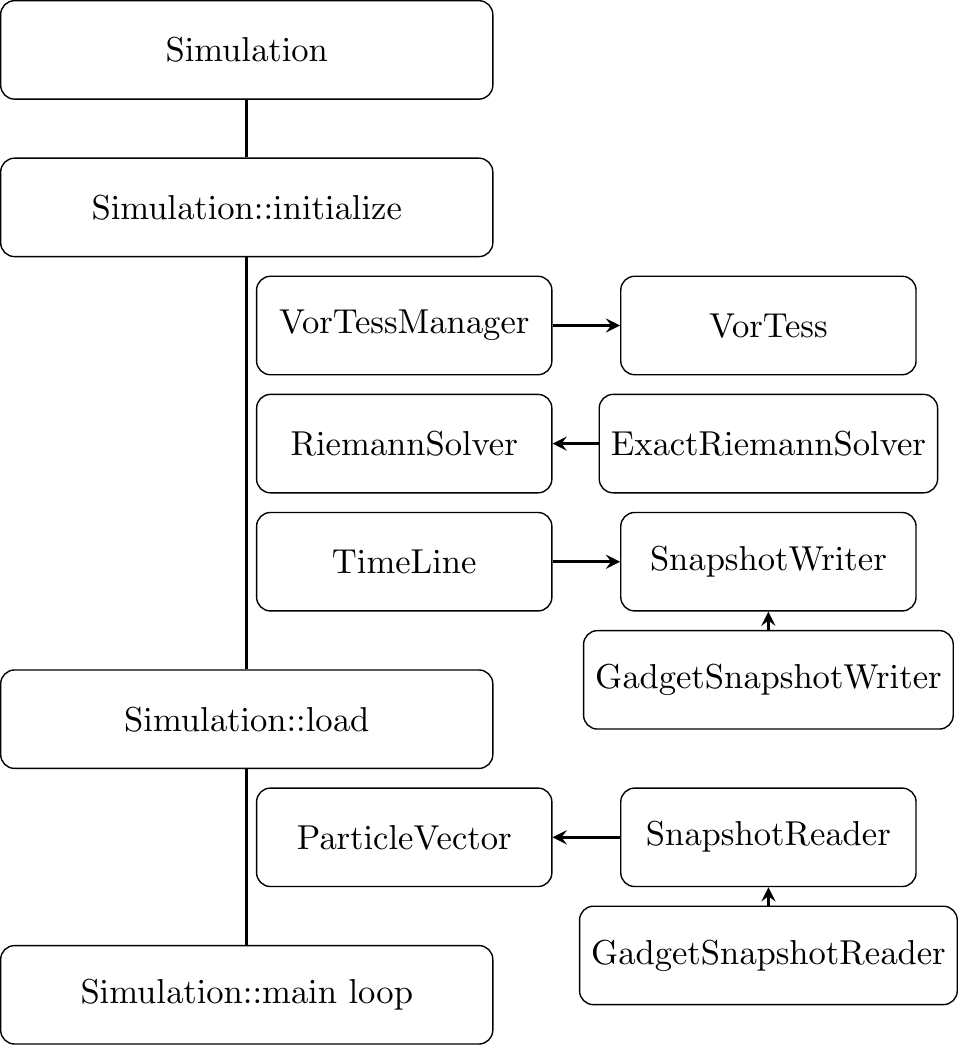}
\caption{General overview of a simulation, using the default Voronoi mesh,
Riemann solver and input/output classes. After the desired implementations of
the main classes have been constructed during an initialization step, the
initial conditions are read in and stored in a \code{ParticleVector} data
structure that will be updated during the main simulation
loop.\label{fig_code_flow_diagram}}
\end{figure}

\begin{figure}
\centering{}
\includegraphics[width=0.5\textwidth]{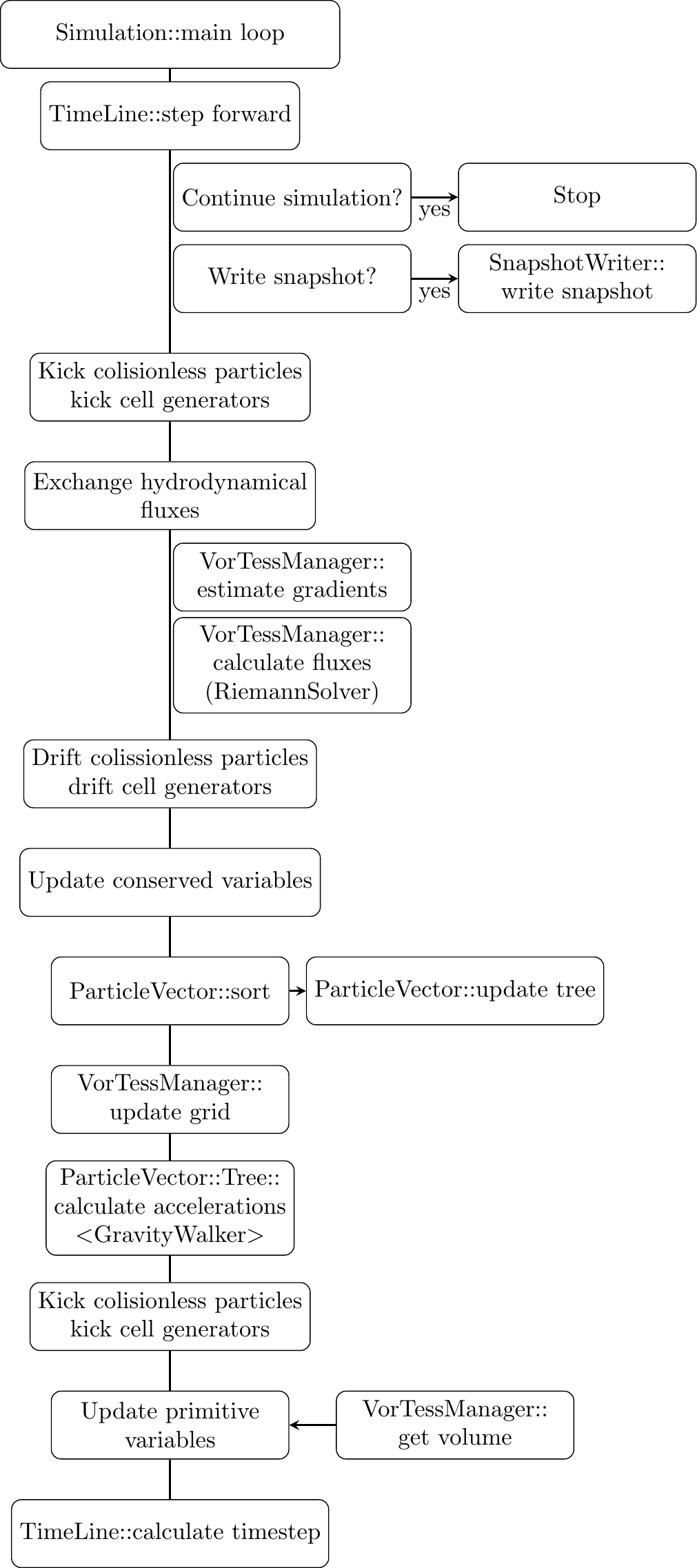}
\caption{Overview of the main simulation loop. We combine a leapfrog integration
for the gravitational force with a simple kick-drift scheme for the
hydrodynamics. Geometrical properties like cell volumes, neighbour relations and
face properties are extracted from the \code{VorTessManager}, gravitational
forces are calculated using the \code{Tree} managed by the
\code{ParticleVector}, using a template \code{GravityWalker}. All time related
stuff is handled by the \code{TimeLine}.\label{fig_main_loop_diagram}}
\end{figure}

\figureref{fig_code_flow_diagram} shows a general overview of a simulation, as
it is implemented in the \code{Simulation} class, introducing the main classes
of interest. \figureref{fig_main_loop_diagram} shows how these different classes
interact during the main simulation loop.

In this section, we describe how various aspects of the algorithm were
implemented in the public version of \textsc{Shadowfax}. We highlight important
abstractions, but also indicate where our current version does not comply with
the high standards formulated above. This is mainly caused by the way in which
the code was developed historically, and changing them will be subject of future
work for versions of the code to come.

We also discuss the parts of the code that are not related to the hydrodynamical
integration, like the gravity calculation, which makes use of advanced C++
language features to increase the readability and abstraction level of the code.
The input and output for initial conditions, parameters, snapshot files and
restart files are discussed as well.

\subsection{Voronoi mesh and hydrodynamics}

We have isolated the Voronoi mesh in a \code{VorTessManager} class, which
encapsulates the entire mesh into a single object. Particles can be added to the
\code{VorTessManager} as a pair consisting of a set of coordinates and an
integer key, so that the properties of the cell or mesh-free particle can then
later be retrieved by means of that same key. The \code{VorTessManager} itself
decides how these properties are calculated. This can be either by using a
Voronoi mesh, by using a fixed Cartesian grid, or even by using a mesh-free
method. The Voronoi mesh could be constructed from the positions of the
particles, but it can also be evolved from the Voronoi mesh from a previous time
step. The public version of \textsc{Shadowfax} for the moment only contains the
ordinary (non-evolved) Voronoi mesh and the fixed Cartesian grid. The evolving
mesh will be added as part of future work (Vandenbroucke \& De Rijcke, in
preparation).

The fluxes are calculated through faces, which could in principle be returned by
the \code{VorTessManager} as \code{VorFace} objects, which contain the surface
area and midpoint of the face, as well as the indices of the two neighbouring
generators. However, due to the historical development of the code and the
complexities involved with parallellizing the Voronoi mesh, the flux calculation
is implemented as a method of the \code{VorTessManager} object for the moment,
with the entire interpolation, integration and flux estimation procedure being
implemented as a method of the \code{VorFace} object.

\subsubsection{Voronoi mesh}

The ordinary Voronoi mesh is contained in a \code{VorTess} object, which
contains an actual list of the faces and of the cells, represented as
\code{VorCell} objects. The \code{VorCell} object contains the necessary methods
to calculate volumes and centroids, and in the current version of the code also
the methods that estimate the gradients and calculate the generator velocities.

When adding a particle to the \code{VorTess}, the particle coordinates are
mapped to the range $[1,2]$ and stored in a \code{VorGen} object, which
represents a mesh generator. This object is stored in a \code{DelTess} object,
which represents the Delaunay triangulation. The \code{DelTess} object also
contains a list of \code{Simplex} objects, which represent the 2D triangles or
3D tetrahedra that constitute the triangulation. Each \code{Simplex} object has
3 or 4 \code{VorGen} objects as vertices (depending on the dimension). During
the incremental construction algorithm, these simplices will change. When all
particles have been added to the \code{VorTess} object, we call an appropriate
method to signal this. At this point, a method of the \code{DelTess} object is
called that will ensure completeness of the Delaunay triangulation (by adding
inactive particles that are neighbour to active particles, and ghost generators
for particles on other MPI processes). When this is done, the actual Voronoi
mesh is constructed by looping over the \code{VorGen} objects stored in the
\code{DelTess} and adding a \code{VorCell} for every \code{VorGen}. For every
neighbour relation between \code{VorGen}s, a \code{VorFace} is created. At the
end of the procedure, we also calculate the volumes, centroids, surface areas
and midpoints of all cells and faces.

\subsubsection{Riemann solvers}

When the flux calculation method of the \code{VorTessManager} object is called,
we pass on a general \code{RiemannSolver} object, which is used to solve the
Riemann problem at the faces, but also to convert between primitive and
conserved variables, and to calculate the actual fluxes. The
\code{RiemannSolver} class itself is just an interface, with the actual
implementation being deferred to child objects. The public version of
\textsc{Shadowfax} implements two Riemann solvers~: an exact Riemann solver
(\code{ExactRiemannSolver} class), and a two rarefaction Riemann solver
(\code{TRRSSolver} class), which is an approximate solver that assumes a wave
structure containing two rarefaction waves.

Which Riemann solver is used, is specified as a run time parameter. To generate
the appropriate \code{RiemannSolver} implementation, a
\code{RiemannSolverFactory} class is used, which implements methods that convert
a string representation of a Riemann solver type to an object. A single
\code{RiemannSolver} object is created during program initialization and used
throughout the simulation, so that the \code{RiemannSolver} can store some
valuable information about the number of Riemann solver evaluations and the
fraction of the run time spent in Riemann solver evaluations.

Since the Riemann solver contains all information about the actual equations
being solved, it is possible to experiment with new equations of state by
writing new \code{RiemannSolver} implementations.

For all tests presented in this paper, we use the exact Riemann solver.

\subsection{Time line and snapshots}

Apart from the spatial discretization, the system of equations is also
discretized in time. We have implemented an individual time stepping scheme, in
which cell time steps are power of 2 subdivisions of the total simulation time.
Fluxes are always exchanged using the smallest time step of the two cells
involved, and the time step of a cell can only increase again if it then becomes
synchronized with other cells on the same time step level. The system is always
evolved using the smallest time step of all cells, but only those cells that are
active at the beginning of the current system time step are actually integrated.

The size of the time step depends on the hydrodynamics and hence should ideally
be set in the \code{RiemannSolver} class. However, for problems involving
gravity, there is also a gravitational time step criterion. Furthermore, the
size of the time step is also governed by a stability parameter (called the
Courant-Friedrichs-Lewy or CFL parameter), which is specified as a run time
parameter for the program. We hence have opted to make the time step calculation
a method of a \code{TimeLine} object, which takes both the hydrodynamical time
step criterion and the gravitational time step criterion into account and
applies the correct CFL parameter. To separate the hydrodynamical aspects, all
hydrodynamically relevant information is encoded in two velocities~: the fluid
velocity and a sound speed, which is calculated by a method of the
\code{RiemannSolver} object.

We also implemented an advanced, tree walk based time step criterion to adapt
the cell time step in the presence of strong shocks. This criterion makes use of
the template tree walks discussed below. However, for simulations involving
regular flow and a high dynamic range, this criterion was found to be very
expensive, so that we disable this criterion by default. This criterion can be
activated by setting the corresponding run time parameter.

The \code{TimeLine} object also keeps track of the system time, and provides
methods to convert floating point time values to an internally used integer
timeline (which is more convenient when using a power of 2 time step hierarchy).
The \code{TimeLine} is also responsible for writing snapshot files.

Snapshots are dumps of the positions and primitive variables of the particles
(along with other relevant quantities) at a particular system time. The time
interval in between snapshots can be specified as a run time parameter and is
stored in the \code{TimeLine}, together with a counter that keeps track of the
number of snapshot files already written. Whenever the system time becomes equal
to or larger than the time for the next snapshot, a signal is sent to an
implementation of the \code{SnapshotWriter} interface, which is responsible for
writing the snapshot. Note that this means that snapshots are not necessarily
written at the exact time requested, if that time does not coincide with an
actual system time. This is a consequence of our strict power of 2 time step
hierarchy and the fact that we do not drift quantities in between system times.
Usually however, the difference will be small. Note also that the use of
individual time steps means that not necessarily all cells will be active at the
time when a snapshot is written. In this case, the primitive variables that were
calculated at the last point in time when a cell was active are used for that
cell. Again, the difference will usually be small.

The \code{SnapshotWriter} interface defines a method that writes the actual
snapshot. Two implementations are provided~: \code{ShadowfaxSnapshotWriter},
which writes snapshots in the historically native \textsc{Shadowfax} format, and
\code{GadgetSnapshotWriter}, which writes snapshots in the same format that is
also used for \textsc{gizmo} and \textsc{swift}, and was one of the possible
snapshot formats for \textsc{Gadget2}. Both formats are based on the HDF5 file
format.

Initial condition files have the same format as snapshot files, and are read in
using the appropriate implementation of the \code{SnapshotReader} interface.
Both the format of the initial conditions as that of the snapshot files can be
specified as a run time parameters for the program, the default format being the
\textsc{Gadget} format. Appropriate objects are generated using the
\code{SnapshotReaderFactory} and \code{SnapshotWriterFactory} classes.

\subsection{Tree algorithms}

Gravitational accelerations are calculated using a Barnes-Hut tree walk
\citep{1986BarnesHut}, using a relative tree opening criterion
\citep{2005Springel}. To this end, a hierarchical octree is constructed~: a
\code{Tree} object. This object holds a pointer to a single implementation of
the \code{Node} interface, which is called the root of the tree. The \code{Node}
interface has two implementations, \code{TreeNode} and \code{Leaf}, which
correspond to respectively a node and a leaf of the octree. Each \code{TreeNode}
has up to 4 or 8 children (depending on the dimension), which themselves are
also \code{Node} implementations, and can be either a \code{TreeNode} or a
\code{Leaf}. Each \code{Leaf} holds a pointer to a single \code{Particle}, which
can be either a \code{GasParticle} or a \code{DMParticle} (see below). In
parallel simulations, there is also a third type of \code{Node}, called a
\code{PseudoNode}, which is used to represent a \code{TreeNode} on another MPI
process.

The tree is constructed using an incremental construction algorithm, and making
use of the close link between the levels of the tree and the levels of a Hilbert
space filling curve \citep{2008Sundar}, the latter also being used for the
domain decomposition \citep{2005Springel}. To this end, we first calculate a
Hilbert key for each particle, using the efficient algorithm of
\citet{2005JinMellorCrummey}. We use a global octree on each process, with nodes
that are entirely on other processes being represented by pseudo nodes.

The gravitational tree walk itself is in essence the same as in
\textsc{Gadget2}, using the same Barnes-Hut tree algorithm with relative opening
criterion and Ewald summation to treat periodic boundaries, but it has been
entirely rewritten to increase readability and code reuse. To this end, we have
defined a general \code{TreeWalker} interface. Every tree walk, be it a tree
walk to obtain gravitational accelerations, or a tree walk to find the
neighbours of a particle within a given radius, is then represented by an
implementation of this interface. The interface itself defines methods that
correspond to the different tasks during a tree walk~: a method to check whether
a node of the tree should be opened or treated approximately, a method called
when a leaf is encountered, and a method that is called when a pseudo node is
encountered and checks whether the tree walk should be continued on another
process. Apart from this, the interface also defines methods that initialize the
variables used during the tree walk, and a method that is called when the tree
walk is finished.

Every possible tree walk is implemented as a single method of the \code{Tree}
class, using the concept of C++ templates. To this end, the tree walk method
takes the name of a \code{TreeWalker} implementation as a template argument, as
well as a list of particles for which the tree walk should be performed. For
every particle in the list, a corresponding \code{TreeWalker} object is created
and used to walk the tree~: we open the root node of the tree and apply the
method that checks if a node should be opened on its children. If a node should
be opened, we continue with its children, if not, we execute all calculations
that are necessary to get the approximate contribution of the entire node to the
tree walk, using the node properties. If a leaf is encountered, the properties
of the corresponding particle are used, and we continue with the next node or
leaf on the same level.

By using class methods to define the different steps in the tree walk, the code
is a lot easier to read, since, for instance, the code that decides if a node
should be opened now is in a separate method that takes a \code{TreeNode} as
single parameter. We do not need to worry about how to implement the tree walk
itself efficiently, only about what happens when a node or leaf is encountered.
By using C++ templates, we limit the overhead usually involved with run time
polymorphism, since the code for a specific type of \code{TreeWalker} is
generated at compile time.

Our approach has some benefits for MPI parallellization as well, since all
explicit communication (and the design of an effective communication scheme) is
limited to the single tree walk method. The only tree walk specific things we
need to do, are deciding if communication is necessary (which is in most cases
similar to deciding whether or not to open a node), deciding what information to
communicate, and how the result of the tree walk on another process should be
communicated back to the original process. To this end, our \code{TreeNode}
interface defines two subclasses, called \code{Export} and \code{Import}.

\subsection{Particles}

The different physical components of the simulation are all represented by
\code{Particle}s, which is an abstract class holding a position and velocity, as
well as an unique identifier (ID), which can be used to trace a single particle
throughout different snapshots. Currently, two subtypes of \code{Particle} are
supported~: \code{GasParticle}s and \code{DMParticle}s, representing
the generators of the Voronoi mesh and a collisionless cold dark matter
component respectively. We have adopted the name \code{GasParticle} instead of
cell to reflect the fact that our hydrodynamical method is just an alternative
for common particle-based hydrodynamical integration schemes. For the
gravitational calculation, the gas is treated as if it consists of particles.

The \code{GasParticle} class extends the particle data with two
\code{StateVector} members, representing the primitive and the conserved
quantities for that cell. It also holds a number of other variables required by
the hydrodynamical integration scheme. The \code{DMParticle} only holds a
particle mass and some auxiliary variables for the gravitational calculation.

The \code{Particle} class itself extends the \code{Hilbert\_Object} interface,
which links a space filling Hilbert key to it. Every class that implements the
\code{Hilbert\_Object} interface can be sorted using an efficient
\code{ParallelSorter}, based on \citet{2010Siebert}. The \code{ParallelSorter}
takes care of the domain decomposition and data size based load-balancing
accross all MPI processes. The domain decomposition is currently only based on
the number of particles, and tries to assign equal numbers to all processes.
This simple approach only works for homogeneous setups with small numbers of
processes, and will be replaced by a more advanced cost-based domain
decompisition in future versions of the code.

\code{Particle}s of all types are stored in a \code{ParticleVector}, which is a
specialized wrapper around two standard C++ \code{vector} objects. The class has
member methods to separately access \code{Particle}s of both types, and is also
responsible for maintaining the \code{Tree} and storing some general information
about the simulation.

\subsection{Units}

We have adopted the strategy used by \textsc{swift}, and force all quantities
that are used as input or output of our program to have units attached to them.
To this end, both snapshot formats have appropriate blocks specifying in what
units the given data are expressed, given in terms of a reference unit system,
which is either the SI or CGS system. If no units are specified for the initial
condition file, SI units are assumed by default. The output units can be chosen
as a run time parameter, as well as the internally used units. Run time
parameters that should have units are assumed to be in internal units, the
default being SI units.

In principle, it does not matter what units are used internally in our code,
although using a system of units in which variables have values close to unity
can be advantageous. Furthermore, all run time log information will be expressed
in internal units, so that it is useful to have some idea of what units are
used.

To simplify working with units, we have implemented a \code{Unit} class. This
class stores the quantity for which the unit is used, expressed as a combination
of the three basic quantities \emph{length}, \emph{mass} and \emph{time}, and
the value of the unit in SI units. \emph{Temperature} and \emph{current} should
be added to the set of basic quantities to be able to express all possible
quantities, but these are not used for the variables that are currently evolved
in \textsc{Shadowfax}.

\code{Unit}s can be multiplied with or divided by one another, yielding a new
\code{Unit}. To make working with quantities in this case possible, we have
adopted some conventions of how quantities should be combined. We also made it
possible to compare different \code{Unit}s, with two \code{Unit}s being
compatible if their quantities are equal. Quantities with compatible
\code{Unit}s can be converted into each other by using a \code{UnitConverter}. A
combination of a length, mass and time unit defines a complete unit system, a
\code{UnitSet}, which is what we specify as a run time parameter for the
program. We currently support three different unit systems~: SI units (m, kg,
s), CGS units (cm, g, s) and galactic units (kpc, M$_\odot{}$, Gyr), which can
be generated using the \code{UnitSetGenerator} class.

Physical constants also have units, but usually their value is fixed. We have
hard coded the values of relevant physical constants (currently only the
gravitational constant $G=6.67408 \times{}
10^{-11}\,\textrm{m}^3\,\textrm{kg}^{-1}\,\textrm{s}^{-2}$) in SI units, and
store their value in internal units in a \code{Physics} object. We have
provided a mechanism to override the physical value of the gravitational
constant at run time for test purposes, but there are no plans to generalize
this approach for other physical constants.

\subsection{Restarting}

Since \textsc{Shadowfax} is meant to be used on small clusters for simulations
that take several days or even weeks to complete, we need to address the
possibility that runs might be interrupted, either by limits on the use of
infrastructure, or by hardware failure. Since the initial condition files and
snapshot files have the same format, it is always possible to restart a run from
the last snapshot. However, since a snapshot is not necessarily written at a
time when all cells are active, this will affect the outcome of the simulation.

We therefore implemented an explicit restart mechanism that dumps the entire
simulation to binary files and then restarts it as if the run never stopped.
This is represented by a \code{RestartFile} object. This object has two template
methods, \code{write} and \code{read}, with several specializations, to write
values in all sorts of formats to the binary file without the need to explicitly
type cast anything. All objects in the simulation that need to be dumped to the
restart file either implement a \code{dump} method, which writes member
variables to the restart file, or are written entirely to the restart file using
one of the \code{write} specializations.

When restarting the run, objects are either created using a special constructor
which initializes values by reading them from the restart file, or are read
entirely from the file using an appropriate \code{read} specialization. To
ensure a proper working of the restart mechanism, we only need to make sure that
values are written and read in the same order.

Since restarting a run is only meant to be done when the run was somehow
interrupted, the \code{RestartFile} also writes a short summary file, containing
compilation, run time and version information about the program that created the
dump. When restarting, this file is read in and the information is compared with
the running program. Only when the same version of the code, with the same
compilation time and the same run time environment is used, do we allow the code
to restart.

\subsection{Future improvements}

The current version of the code still contains traces of the way in which it was
originally developed, that do not comply with our strict design goals. However,
refactoring the code to eliminate these infers a major update, and is postponed
to a future version of the code. We give a brief overview below.

\begin{itemize}
\item \code{VorFace} still contains the entire gradient reconstruction and
prediction step, as well as the flux calculation. These should be isolated into
a new class that use the properties of the \code{VorFace}, but can be decoupled
from its geometric meaning.
\item Similarly, \code{VorCell} still contains the gradient estimation and the
generator velocity calculation. These should be calculated inside a new class
that uses the \code{VorCell} properties.
\item Currently, only the time step tree walk and the gravitational tree walk
use a template \code{TreeWalker}, the neighbour search is still hardcoded in the
\code{Tree} class.
\item The MPI communication for the cells is strongly coupled to ghost
\code{VorGen}s stored in the \code{DelTess}. As a result, all cell communication
has to go through the \code{VorTessManager}, which is not ideal.
\item The domain decomposition does not take into account the effective
computational cost on an MPI process, but is only based on particle number.
\item The \code{FixedGrid} only works if the particles have specific positions
and do not change their positions. This means the \code{VorTessManager} should
have control over the positions of the particles, and no other class.
\end{itemize}

\section{Setup and analysis}

Apart from the main simulation program, the public release of \textsc{Shadowfax}
also includes two auxiliary programs that can be used to generate initial
condition files, and to convert snapshot files to VTK files that can be easily
visualized using common visualization packages. These are discussed below.

\subsection{Initial condition generation}

Since the default file format for \textsc{Shadowfax} is the same as for a number
of important other astrophysical codes, it is possible to use software written
for those codes to generate initial conditions for \textsc{Shadowfax}, or to
analyse \textsc{Shadowfax} results. We however also provide our own initial
condition generating software, which is a part of \textsc{Shadowfax}. It is also
worth noting that HDF5 has a user-friendly Python interface,
\code{h5py}\footnote{\url{http://www.h5py.org}}.

Our own program is called \code{icmakerXd}, (with \code{X} being 2 or 3), and is
based on the initial condition mechanism implemented in the AMR code
\textsc{ramses}\footnote{ascl:1011.007} \citep{2002Teyssier}. The simulation box
is divided into a number of geometrical \emph{regions}, each having a
geometrical shape and values for the primitive hydrodynamical variables inside
that region. Multiple regions can overlap, in this case the values for the
region that was last added are used in the overlap region.

The shape of the region is set by defining the position of the origin
$\boldsymbol{o} = (o_x, o_y, o_z)$ of the region, and 3 widths, $w_x$, $w_y$,
$w_z$, together with an exponent $e$. The latter is used to determine if a point
$\boldsymbol{p} = (p_x, p_y, p_z)$ lies inside the region. For this, the
inequality
\begin{equation}
\left[ \left( 2\frac{o_x - p_x}{w_x} \right)^e + \left( 2\frac{o_y - p_x}{w_y}
\right)^e + \left( 2\frac{o_z - p_z}{w_z} \right)^e \right]^\frac{1}{e} \leq{} 1
\end{equation}
needs to hold. For 2D setups, we only have 2 widths and the last term drops out
of this inequality.

We have extended the \textsc{ramses} scheme and do not only allow constant
values for the primitive variables inside a single region, but also more complex
expressions, containing mathematical operations ($+$, $-$, $\times{}$, $/$),
basic mathematical functions ($\cos{}$, $\sin{}$...), mathematical constants
($\pi{}$), and even coordinate expressions ($x$, $y$, $z$ and $r$). Support for
these expressions is provided by \textsc{Boost Spirit}\footnote{\url{
http://www.boost.org/doc/libs/release/libs/spirit/}}.

\code{icmakerXd} has support for regular Cartesian setups, but also for random
unstructured grids, whereby grid generators are sampled according to the
hydrodynamical density of the regions, using rejection sampling. To this end,
the (potentially) complex density profiles inside the different regions are
numerically integrated to obtain weighing factors. Allowing complex expressions
makes this computationally expensive, but makes the program a powerful tool for
initial condition generation.

To smoothen out Poisson noise in random generator setups, we apply 10 iterations
of Lloyd's algorithm \citep{1982Lloyd}. This yields a more regular initial mesh.

\subsection{Visualization}

Visualizing a Voronoi mesh is not so straightforward, especially in the case of
a 3D mesh. Furthermore, the vertices of the Voronoi mesh are not written to the
snapshot files, so that the mesh needs to be recalculated if we want to
visualize the mesh corresponding to some snapshot file. This requires a Voronoi
construction algorithm, and hence we have written an auxiliary program, called
\code{vtkmakerXd} that converts a regular snapshot to a dump of the Voronoi
mesh, in the VTK file format\footnote{\url{http://www.vtk.org}}. This file
format can be read by software that makes use of the powerful Visualization
Toolkit (VTK), the most commonly used examples being
Paraview\footnote{\url{http://www.paraview.org/}} and
VisIt\footnote{\url{https://wci.llnl.gov/simulation/computer-codes/visit}}.

We have also written a plugin for VisIt that reads in the default
(\textsc{Gadget}) snapshot format. It should also be possible to use the
eXtensible Data Model and Format (XDMF)\footnote{\url{http://www.xdmf.org}} and
the XDMF plugins for VisIt and Paraview, an approach taken by e.g.
\textsc{swift}. However, the current version of \textsc{Shadowfax} does not
write the necessary XDMF file, so that it needs to be manually created.

\section{Basic tests}

To validate the code, we have used it to evaluate a number of test problems.
These have been gathered into a \code{testsuite}, together with the necessary
files to create the initial conditions, run the simulations, and analyze the
results. Where possible, we have provided analytic solutions to compare with.

These tests are meant to be run as a general code check after every significant
change in the code, and for this reason use relatively low resolution grids. We
therefore will not focus on obtaining the best possible accuracy. Rather, we
will focus on accuracy when we compare \textsc{Shadowfax} with other publicly
available codes in a later section.

In this section, we describe the physical test problems currently in the
\code{testsuite} and discuss their results. We do not describe tests that are
used to verify the proper working of the program itself, like the
\code{restarttest}, which checks if the program correctly restarts from restart
files.

\subsection{Spherical overdensity test}

This test problem consists of a uniform box with unit length, in which a fluid
with density $0.125$ and pressure $0.1$ is in rest. In the center of the box, a
spherical overdense region with density $1$, pressure $1$, and radius $0.25$
is inserted. This test corresponds to one of the Riemann solver tests in
\citet{2009Toro}, but generalized to 2 or 3 dimensions to test geometrical
aspects of the code as well.

\begin{figure*}
\centering{}
\includegraphics[width=\textwidth]{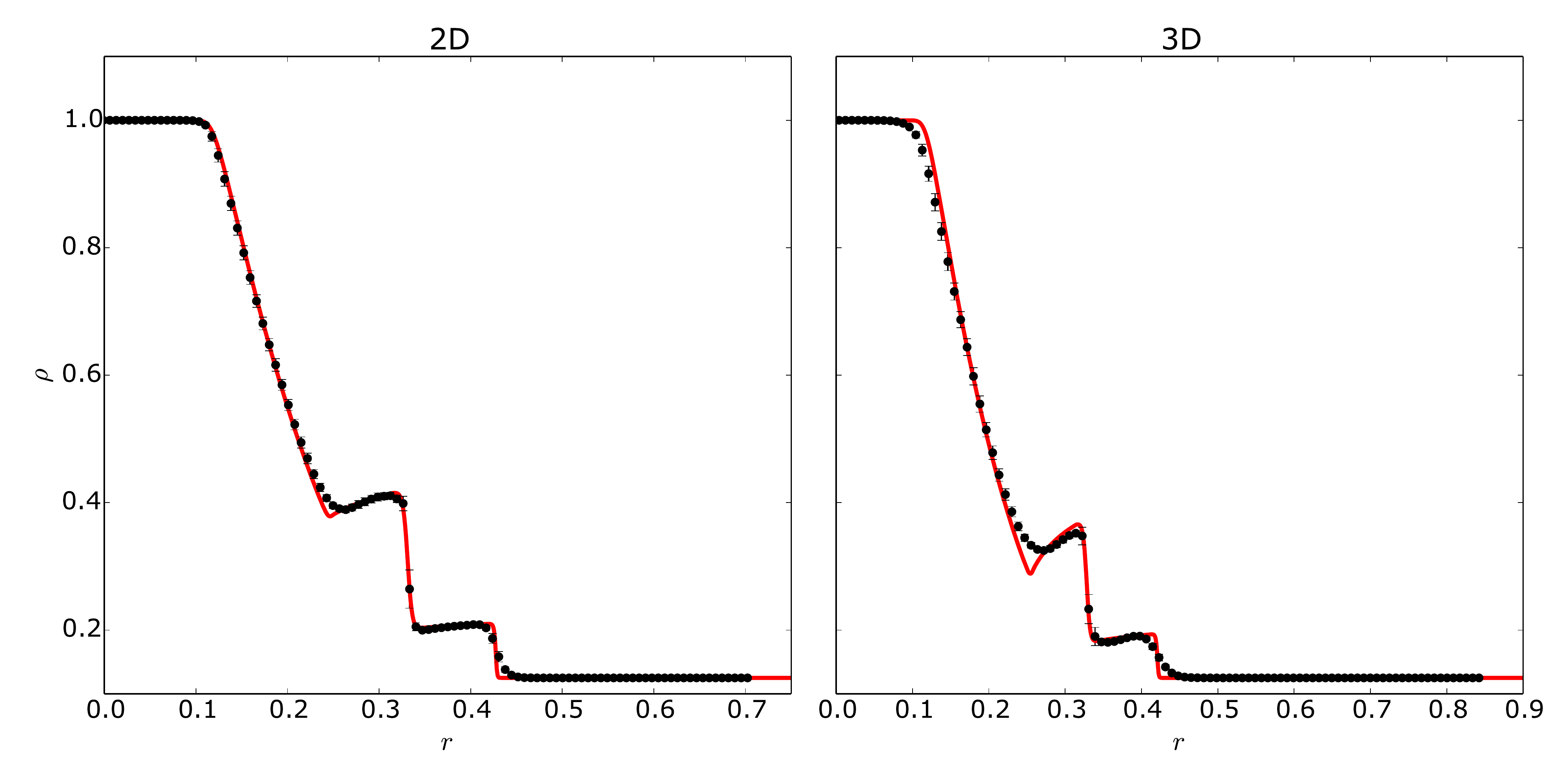}
\caption{Radial density profile for the spherical overdensity test at time
$t=0.1$. \emph{Left:} 2D result, using 10,000 uniformly sampled cells.
\emph{Right:} 3D result, using 100,000 uniformly sampled cells. The black dots
are the simulation results, the red line corresponds to the solution of the high
resolution 1D equivalent problem. To limit the number of data points, the
simulation results have been binned, the standard deviation of the density
within the bins is indicated by the error flags.\label{fig_overdensity}}
\end{figure*}

The solution of this problem consists of an inward travelling rarefaction wave,
a central contact discontinuity, and an outward travelling shock wave. The
simulation results in 2 and 3 dimensions  at time $t=0.1$ are shown in
\figureref{fig_overdensity}, together with a high resolution result for the
equivalent 1D problem obtained using a finite volume method on a fixed 1D grid.
The different features of the solution are clearly resolved.

\subsection{Gresho vortex}

For this test, a vortex in hydrostatic equilibrium is evolved for some time to
check the local conservation of angular momentum. Inside a box with unit length
and constant density $1$, a 2D azimuthal velocity profile of the form
\citep{2010Springel}
\begin{equation}
v_\phi{} (r) = \begin{cases}
5r & 0 \leq{} r < 0.2\\
2 - 5r & 0.2 \leq{} r < 0.4\\
0 & 0.4 \leq{} r
\end{cases}
\end{equation}
is balanced by a pressure profile of the form
\begin{equation}
p(r) = \begin{cases}
5 + \frac{25}{2}r^2 & 0 \leq{} r < 0.2 \\
9 + \frac{25}{2}r^2 - 20r + 4\log{}\left(\frac{r}{0.2}\right) & 0.2 \leq{} r <
0.4 \\
3 + 4\log{}(2) & 0.4 \leq{} r.
\end{cases}
\end{equation}

\begin{figure*}
\centering{}
\includegraphics[width=\textwidth]{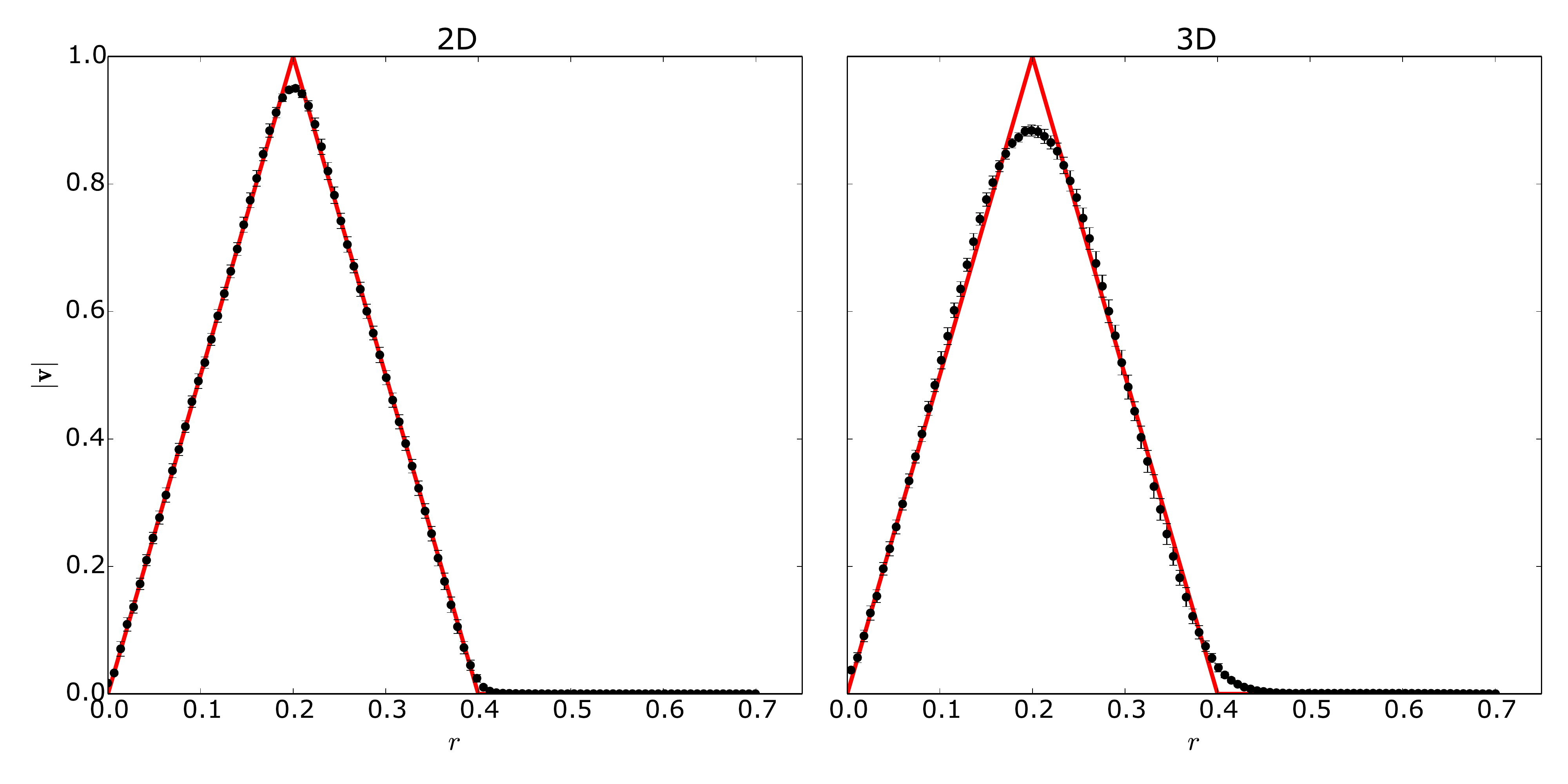}
\caption{Radial velocity profile for the Gresho vortex test at time $t=3$.
Both the 2D and the 3D simulation use 10,000 uniformly sampled cells. The black
dots represent the binned simulation results, with the error flags indicating
the standard deviation on the values within the bins. The red line represents
the initial velocity profile, which should remain constant.\label{fig_gresho}}
\end{figure*}

In 3D, we use the same profile and expand it to cylinders in the third
dimension, in a box of size $1\times{}1\times{}1/3$. We evolve the setup
to time $t=3$, and compare the velocity and pressure profile with the initial
profiles. The results are shown in \figureref{fig_gresho}.

Both the 2D and 3D test conserve the hydrostatic equilibrium relatively well
over a long time scale. This means local angular momentum is conserved to a
reasonable degree.

\subsection{Sedov-Taylor blast wave}

This problem is meant to test the limits of the code, both of the Riemann
solver, the mesh regularization algorithm, and the time step criterion. It
consists of a box with unit length in which a cold medium with density $1$ and
pressure $10^{-6}$ is in rest. In the central cell, we set the pressure to a
much higher value, which corresponds to an energy input of $1$, so that a
strong explosion is initiated. To accurately capture the explosion, it is
important that
\begin{itemize}
\item the central cells are kept regular at the start of the simulation
\item the cells surrounding the center are given small enough individual time
steps to be active when the shock arrives
\item the Riemann solver is able to handle vacuum generating conditions to
correctly estimate fluxes around the central cell
\end{itemize}

\begin{figure*}
\centering{}
\includegraphics[width=\textwidth]{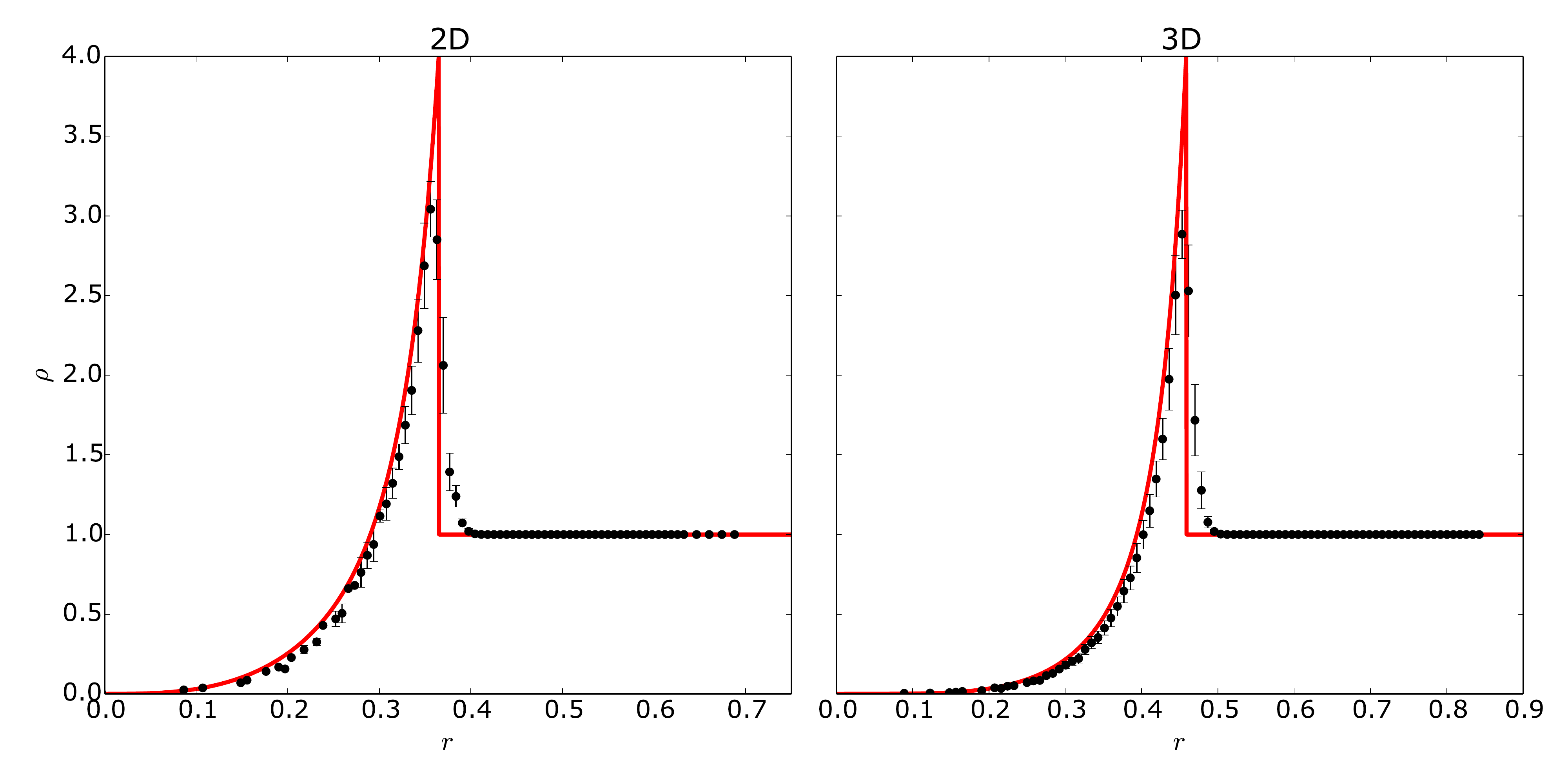}
\caption{Density profile for the Sedov-Taylor blast wave test at time $t=0.1$.
{Left:} 2D version, using an initially Cartesian grid with $45\times{}45$ cells.
{Right:} 3D version, using an initially Cartesian grid with
$45\times{}45\times{}45$ cells. The black dots represent the binned simulation
results, with the error flags indicating the standard deviation of the density
values within the bins. The full red line is the analytical solution of
\citet{1977Sedov}.\label{fig_sedov}}
\end{figure*}

The resulting shock profile is self-similar and has an analytic solution
\citep{1977Sedov}, which is shown together with the simulation results in
\figureref{fig_sedov}. Again, the simulation results are in line with the
analytic solution, and the shock is well-resolved.

\subsection{N-body test}

This test is used to validate the gravitational part of the code, and uses cold
dark matter instead of gas. Since gravity is not guaranteed to work in 2D, this
problem is only provided in 3D.

A Plummer sphere \citep{1911Plummer} with mass $1000$ and scale parameter
$1$ is initiated inside a large box (the actual box is irrelevant for this
specific problem, but \textsc{Shadowfax} always requires a simulation box to be
present). The velocities of the particles are chosen so that the entire problem
is independent of time. For convenience, we set the gravitational constant $G=1$
for this problem. We adopt a gravitational softening length of $0.03$. 

\begin{figure}
\centering{}
\includegraphics[width=0.5\textwidth]{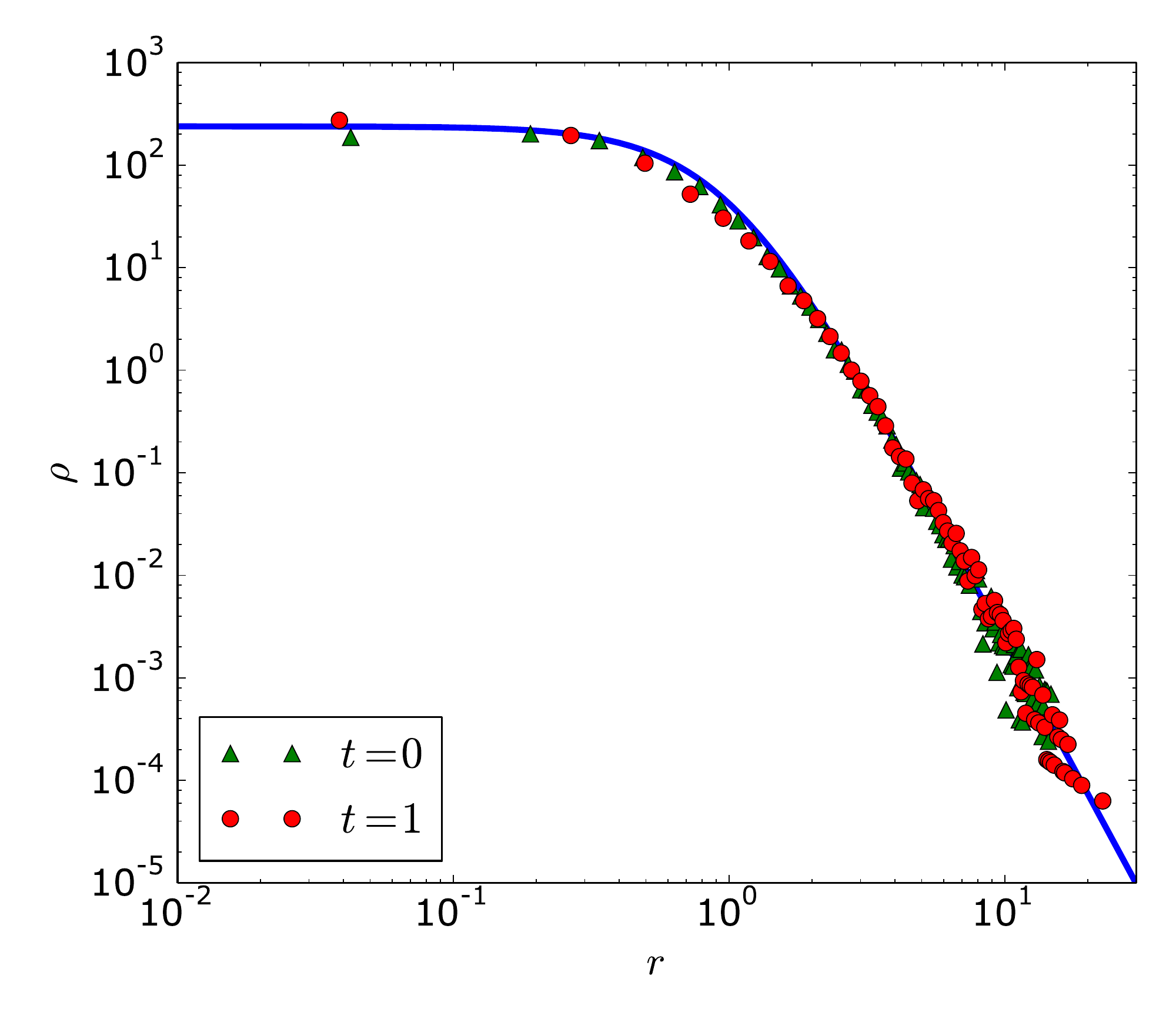}
\caption{Density profile of the N-body test at the start and end of a simulation
using 10,648 cold dark matter particles. The density is calculated by summing
the masses of all particles within spherical shells. The dots and triangles
represent simulation results, the full blue line is the theoretical Plummer
density profile from which the initial condition is
sampled.\label{fig_nbody_profile}}
\end{figure}

\begin{figure}
\centering{}
\includegraphics[width=0.5\textwidth]{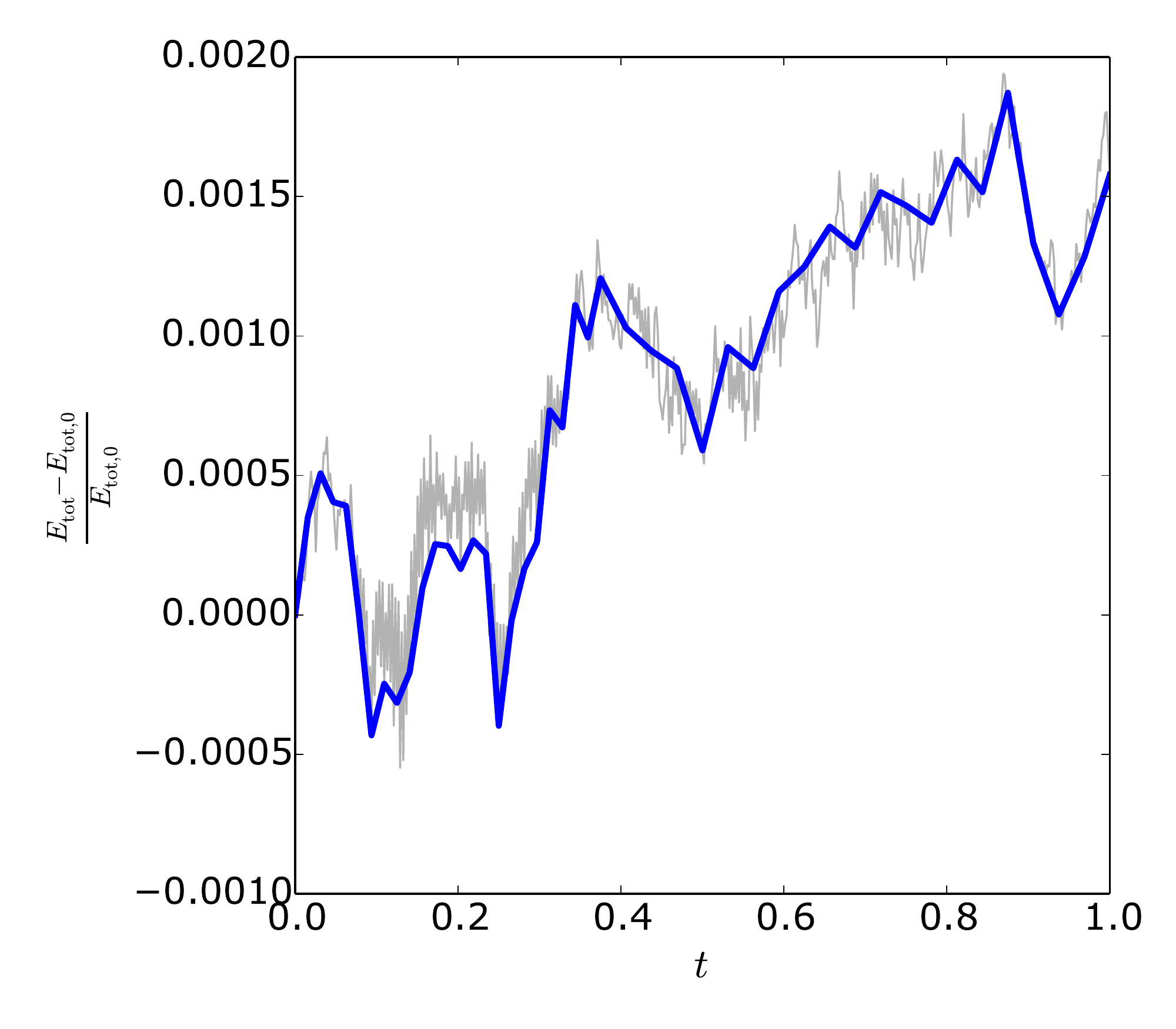}
\caption{The relative energy error of the N-body test as a function of
time. The gray line shows the energy error at all times, the blue line shows the
energy error at the times when all particles are
active.\label{fig_nbody_energy}}
\end{figure}

In \figureref{fig_nbody_profile}, we show the initial density profile, and the
density profile at time $1$, when the system has evolved for $\sim{}10$
dynamical times. \figureref{fig_nbody_energy} shows the relative error on the
total energy. We see that the system remains stable and that the total energy is
quite accurately preserved.

\subsection{Evrard collapse}

This problem tests the coupling between hydrodynamics and gravity, and consists
of a self-gravitating gas cloud with mass $1$ and radius $1$, with a density
profile of the form \citep{2010Springel}
\begin{equation}
\rho{}(r) = \begin{cases}
\frac{1}{2\pi{}(r + 0.001)} & r \leq{} 1 \\
0 & 1 < r.
\end{cases}
\end{equation}
The cloud is initially at rest and has a very low pressure profile of the form
\begin{equation}
p(r) = \begin{cases}
\frac{0.05}{3\pi{}(r + 0.001)} & r \leq{} 1 \\
0 & 1 < r,
\end{cases}
\end{equation}
corresponding to a low constant thermal energy. We set the gravitational
constant $G=1$ for this test and only consider a 3D version. The softening
length is set to $0.003$.

The vacuum boundary of the cloud poses a challenge for our finite volume method,
since small numerical errors on the fluxes might easily cause non-physical
negative masses and energies in the empty cells that surround the cloud. It
forms a good test for our flux limiter and the overall stability of the code,
while at the same time not contributing significantly to the result.

\begin{figure}
\centering{}
\includegraphics[width=0.5\textwidth]{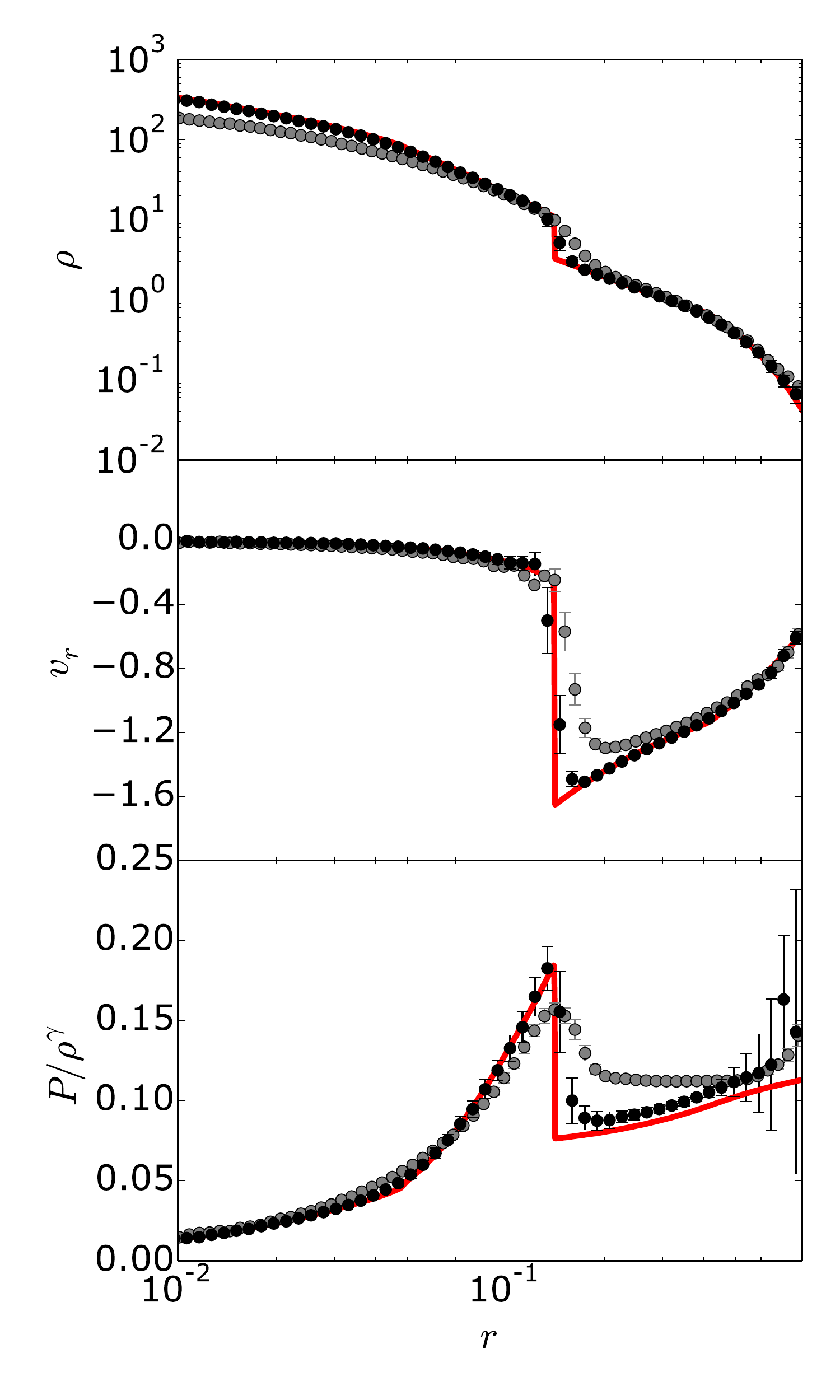}
\caption{Profiles of the Evrard collapse test at time $t=0.81$, when the central
virializing shock has formed, but has not yet reached the boundary of the cloud.
Top: density profile, middle: radial velocity profile, bottom: entropy profile.
The simulation used 20,000 cells in total, some of which are vacuum cells
surrounding the cloud. The black dots represent the binned simulation results,
with the error bars indicating the standard deviation on the density values
within the bins. The red line is the solution of the equivalent 1D problem. The
gray points are the solution for the same setup, but using the SPH code
\textsc{Gadget2}.\label{fig_evrard_profile}}
\end{figure}

\begin{figure}
\centering{}
\includegraphics[width=0.5\textwidth]{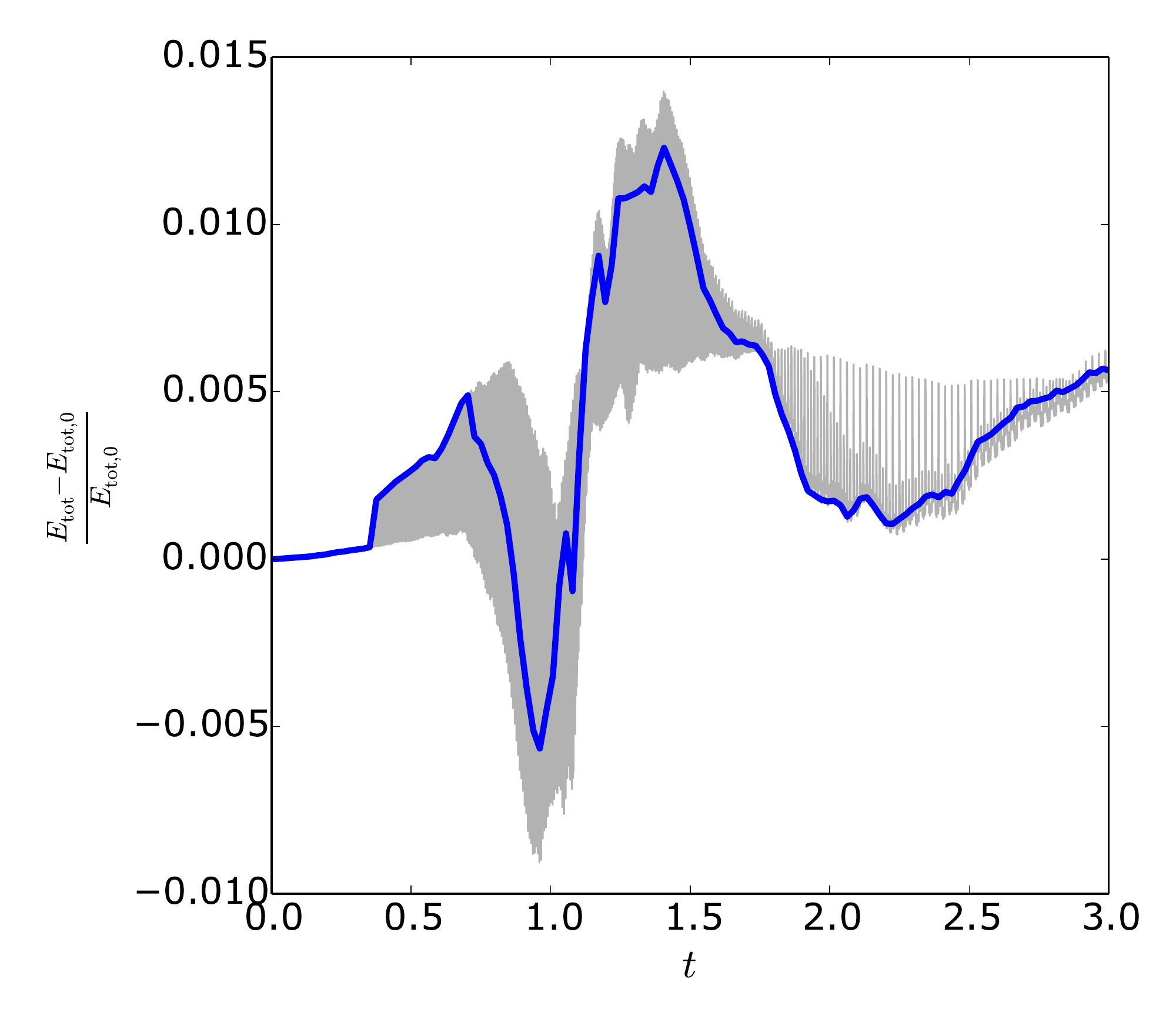}
\caption{The relative energy error for the Evrard collapse test as a function of
time. The gray line represents the energy error at all times, the blue line
represents the energy error at times when all cells are
active.\label{fig_evrard_energy}}
\end{figure}

The density, radial velocity and entropy profiles of the simulation at
time $t=0.81$ are shown in \figureref{fig_evrard_profile}, together with a high
resolution result for the equivalent 1D problem obtained using a finite volume
method on a fixed 1D grid, and the solution of the same setup using the SPH code
\textsc{Gadget2}. The relative error on the total energy is shown in
\figureref{fig_evrard_energy}. The virializing shock is clearly resolved and
travels at the expected speed. We see that total energy fluctuates when the
shock is formed and stabilizes after the system has virialized, which indicates
that the gravitational correction terms in our scheme are not entirely effective
at the current resolution. Better energy conservation can be obtained by using
more resolution, but this is too computationally expensive for the testsuite.

\section{Convergence rate}

The MUSCL-Hancock finite volume method implemented in \textsc{Shadowfax} is
nominally second order in both space and time. However, due to the use of slope
limiters and flux limiters, the actual order of the method can be lower in the
presence of strong discontinuities. To test this, we study the convergence rate
as a function of the number of cells for a number of different tests: a one
dimensional smooth travelling sound wave, a one dimensional shock tube and a two
dimensional vortex.

\subsection{Sound wave}

The initial condition for this test consists of a periodic cubic box with unit
side containing a sound wave with a small amplitude $A=10^{-6}$. The density is
given by
\begin{equation}
\rho{}(x) = 1 + A\sin{2\pi{}x},
\end{equation}
the pressure is $p(x) = \rho{}(x)^\gamma{}/\gamma{}$, with $\gamma{}=5/3$, and
the velocity is zero everywhere. The wave travels with the sound speed, $c_s=1$
for this particular choice of pressure.

Although this problem is formally one dimensional, we study it in 3D, using
$N^3$ comoving cells, with $N$ the number of cells in one dimension. We use both
an initial Cartesian grid (which remains Cartesian throughout the simulation as
the fluid velocity is zero), and a random uniform grid. We expect the
convergence to be better in the former case, as the Cartesian solution
corresponds to a combination of $N^2$ 1D solutions. For a random uniform grid,
the cells are irregularly shaped, and the problem is effectively
multi-dimensional.

We study the convergence rate by comparing the solution $\rho{}_i$ at $t=1$,
when the wave has travelled for one box length, with the initial condition
$\rho{}(x_i)$ at $t=0$. To this end, we calculate the L1 norm
\begin{equation}
\text{L1} = \frac{1}{N^3} \sum_i | \rho{}_i - \rho{}(x_i) |,
\end{equation}
for all cells $i$.

\begin{figure}
\includegraphics[width=0.5\textwidth]{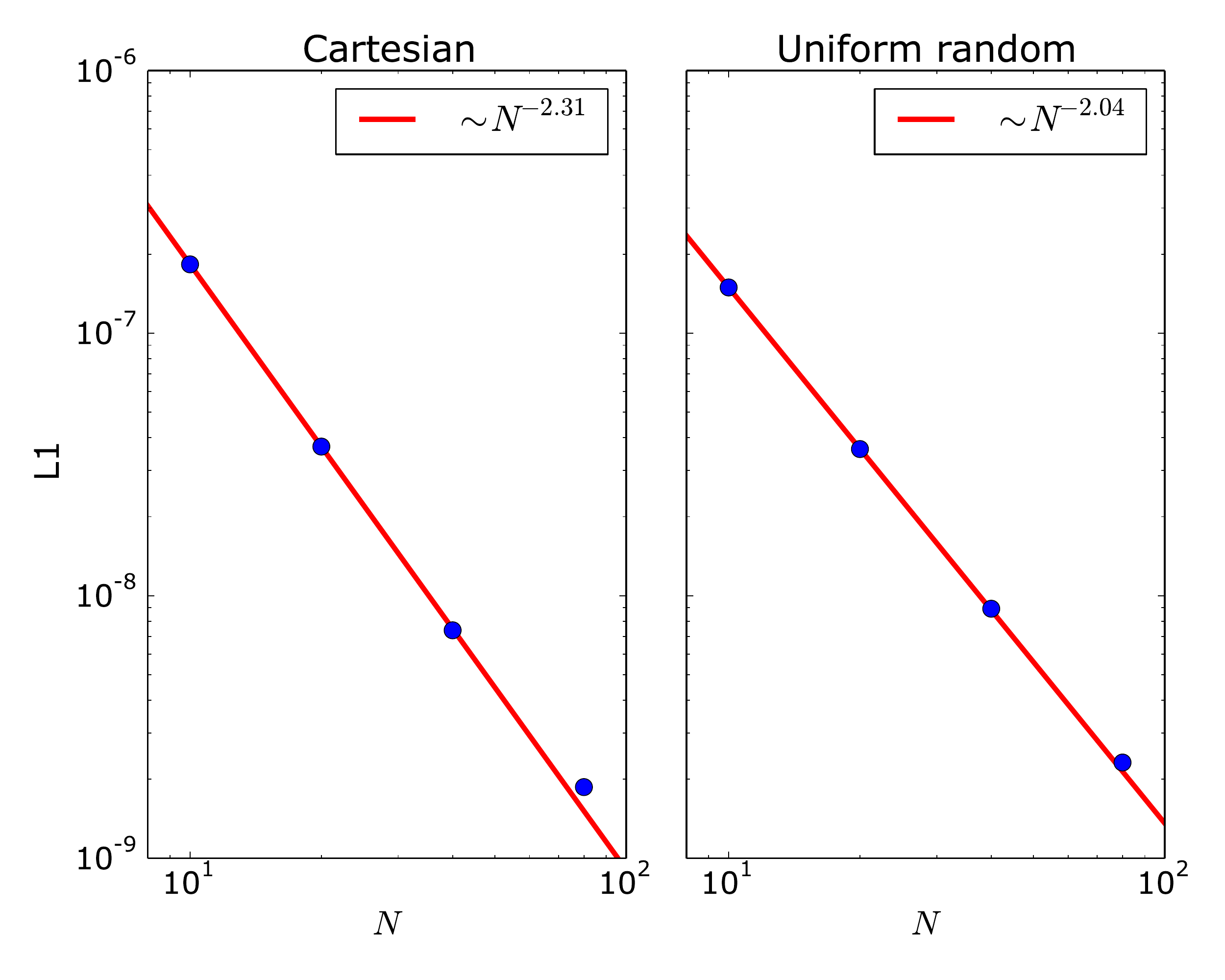}
\caption{L1 norm as a function of 1D cell number for a travelling sound wave in
3D. The blue dots are the simulation results, the red line is a least squares
fit.\label{fig_L1_wave}}
\end{figure}

\figureref{fig_L1_wave} shows the L1 norm as a function of the equivalent 1D
cell number $N$. Both the Cartesian and the random uniform grid show close to
$N^{-2}$ scaling, indicating that the method is indeed second order in space and
time.

\subsection{Sod shock}

A more demanding test problem is the Sod shock test, which consists of a
reflective cubic box with unit length in which the density is given by
\begin{equation}
\rho{}(x) = \begin{cases}
1 & x < 0.5 \\
0.25 & 0.5 \leq{} x,
\end{cases}
\end{equation}
and the pressure by
\begin{equation}
\rho{}(x) = \begin{cases}
1 & x < 0.5 \\
0.1795 & 0.5 \leq{} x.
\end{cases}
\end{equation}
The initial velocity is zero everywhere. This problem is the 1D equivalent of
the spherical overdensity discussed as part of the testsuite, and its solution
has the same characteristic wave components, including a contact discontinuity
and a shock wave, for which we expect the convergence rate of our scheme to be
worse than second order.

We evolve the test to $t=0.12$, and calculate the L1 norm in the same way as
above, but with $\rho{}(x_i)$ now given by the exact solution to the equivalent
Riemann problem at $t=0.12$, which can be found using an exact Riemann solver.

\begin{figure}
\includegraphics[width=0.5\textwidth]{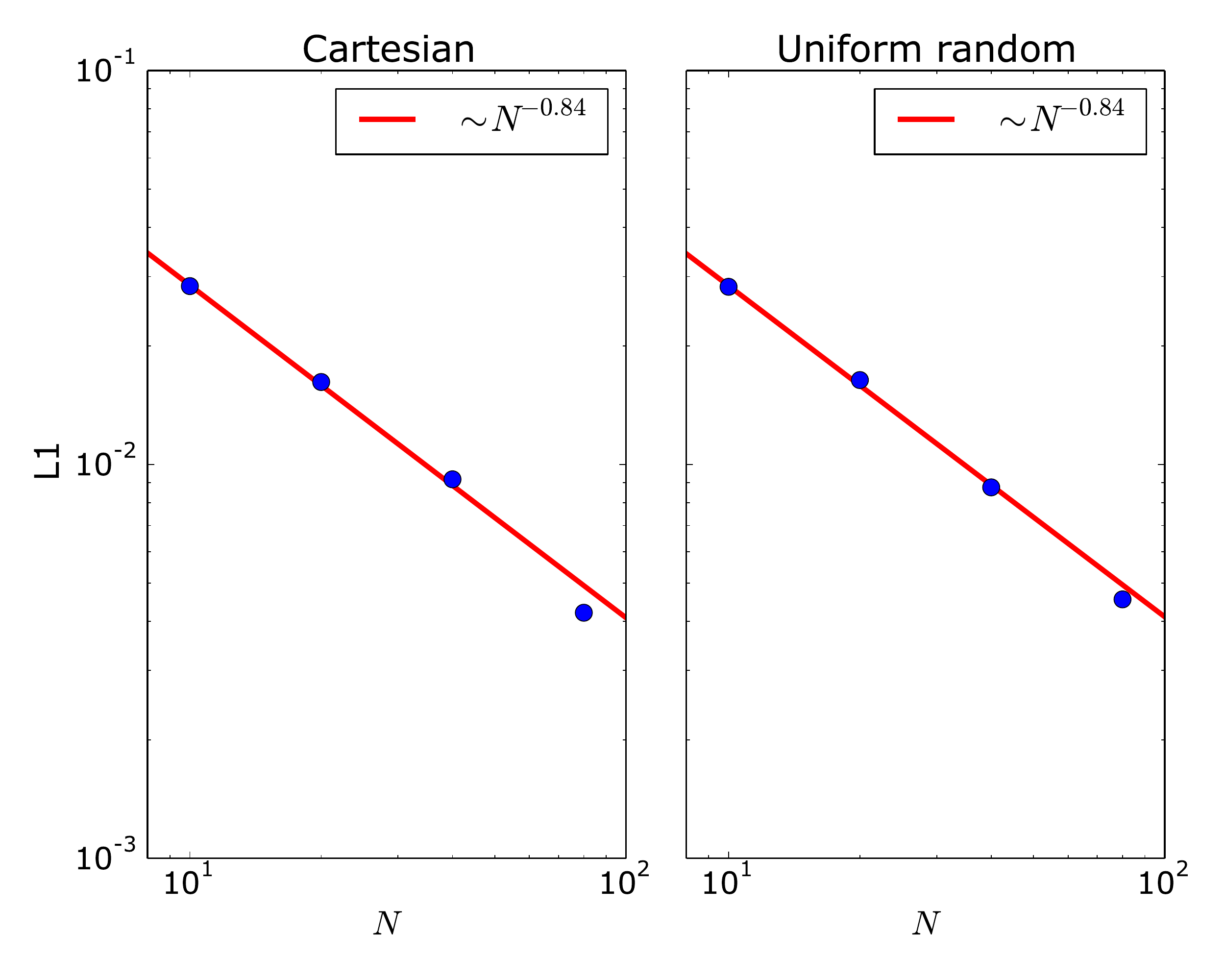}
\caption{L1 norm as a function of 1D cell number for a Sod shock test in 3D. The
blue dots are the simulation results, the red line is a least squares
fit.\label{fig_L1_sod}}
\end{figure}

\figureref{fig_L1_sod} shows the L1 norm as a function of the equivalent 1D cell
number $N$, again for both a Cartesian initial grid and a random uniform initial
grid. In this case, convergence is clearly worse than $N^{-2}$, and is not even
first order in space and time. There no longer is a clear difference between the
Cartesian grid and the uniform random grid.

\subsection{Gresho vortex}

To study the convergence in a manifestly multi-dimensional problem, we also
consider the 2D Gresho vortex, which we already encountered as part of the
testsuite. We run four simulations with respectively 1,000, 10,000, 50,000 and
100,000 random uniform cells until $t=3$, and calculate the L1 norm for the
azimuthal velocities $v_{\phi{},i}$ using the time independent initial condition
$v_{\phi{}}(r)$ given above:
\begin{equation}
\text{L1} = \frac{1}{N} \sum_i | v_{\phi{}, i} - v_\phi{}(r_i) |.
\end{equation}

\begin{figure}
\includegraphics[width=0.5\textwidth]{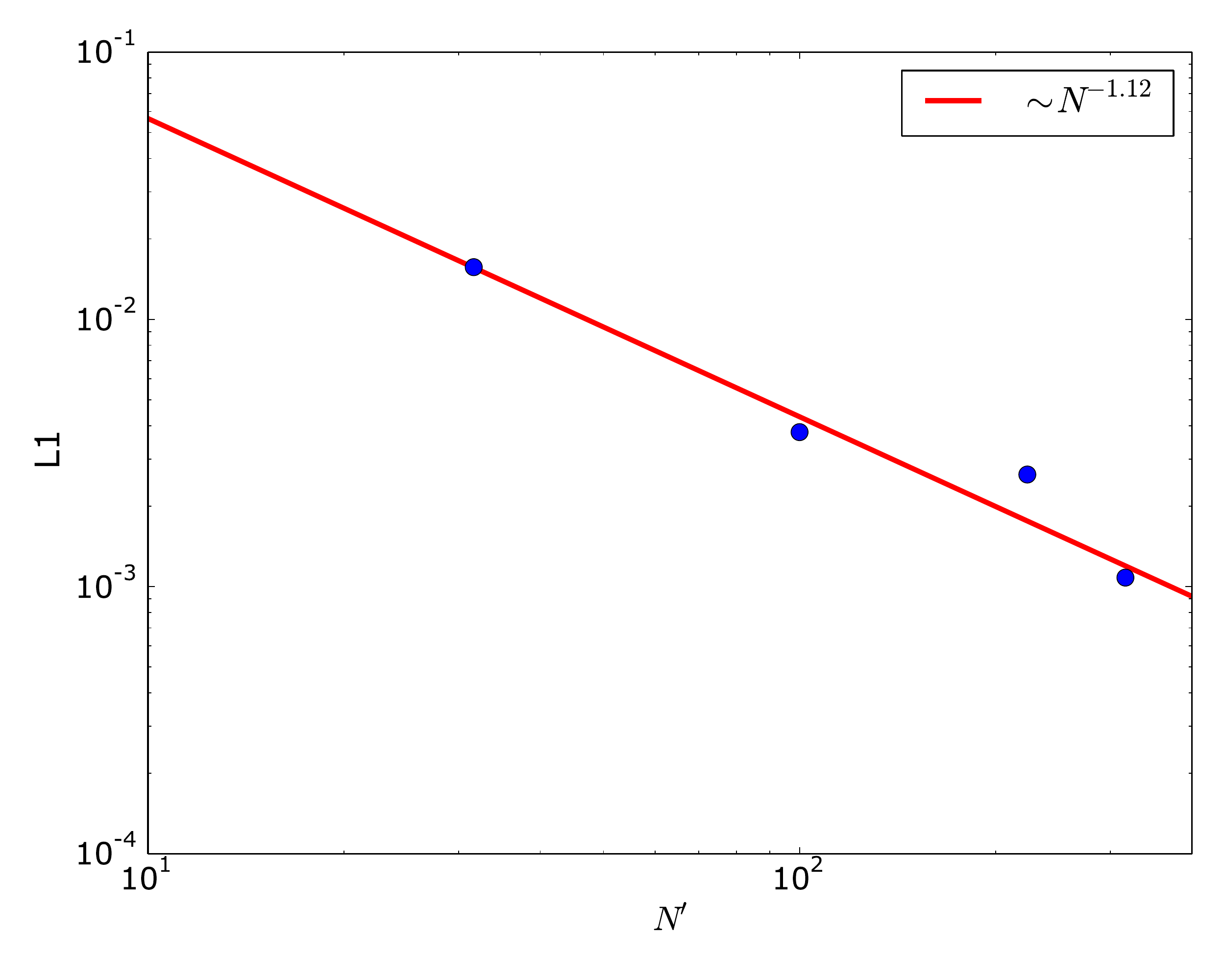}
\caption{L1 norm as a function of 1D cell number for a 2D Gresho vortex test.
The blue dots represent simulation values, the red line is a least squares
fit.\label{fig_L1_gresho}}
\end{figure}

In \figureref{fig_L1_gresho}, we show the L1 norm as a function of the 1D
equivalent cell number $N'=\sqrt{N}$. Again, convergence is not second order,
but it is clearly better than in the case of a strong shock, as the Gresho
vortex is overall more smooth.

\section{Performance}

Modern high performance computing systems consist of large numbers of computing
nodes, each of which can have multiple CPUs, which are made up of multiple
computing cores. These systems are hence highly parallel and employing their
full power requires algorithms that can exploit this parallelism. Compared to
the available computing power, memory is relatively scarce on most system, so
that it is also important to minimize the memory imprint of an application.

The current version of \textsc{Shadowfax} was not optimized for usage on high
performance systems, but some basic MPI communication instructions were added to
make it run on distributed memory systems. To fully exploit the power of modern
architectures, a hybrid algorithm that combines distributed memory parallelism
with shared memory parallelism is needed, as well as a better representation of
the Voronoi mesh that depends less on a globally constructed mesh. This requires
a thorough refactoring of the code, which is left for future versions.

In this section, we discuss the strong and weak scaling of the current version
of the code, the contribution of the various components of the code to the total
run time, and their memory imprint. For all these tests, we use the Evrard
collapse test discussed above, which uses both the hydrodynamical solver and the
N-body solver. Since this setup is very inhomogeneous, our simple domain
decomposition leads to serious load-imbalances. For comparison, we also show
scaling results for the more homogeneous spherical overdensity test.

\subsection{Strong scaling}

Strong scaling measures the decrease in total simulation run time when running
the same simulation on an increasingly large number of processes. Ideally,
doubling the amount of processes should half the total run time, but due to
communication overhead this can never be achieved. As communication scales with
the surface area of the different computational domains in the simulation, while
the computation time per domain scales with the domain volume, we expect the
relative contribution of communication to the total run time to increase with
increasing process number. This poses a natural limit on the speed up that can
be achieved for a given problem size.

\begin{figure}
\centering{}
\includegraphics[width=0.5\textwidth]{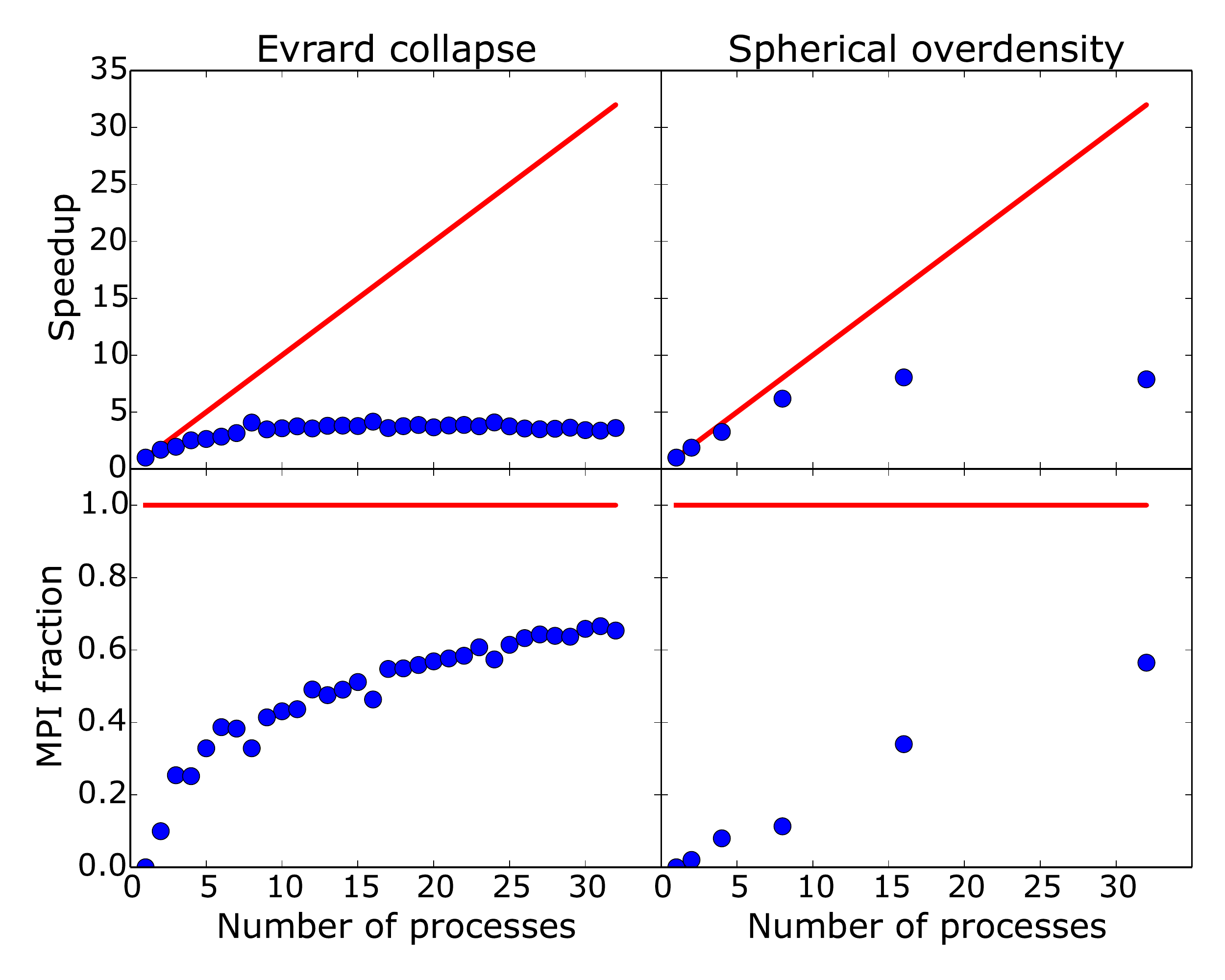}
\caption{Speedup and MPI fraction for two different tests from the
testsuite under strong scaling.\label{fig_scaling_strong}}
\end{figure}

\figureref{fig_scaling_strong} shows the strong scaling for two of the test
problems that are part of the testsuite: the 3D Evrard collapse and the 3D
spherical overdensity. The tests were run on a single node of our local
computing cluster, consisting of 4 2.7 GHz Intel Xeon CPUs with 8 cores each,
using version 4.8.4 of the GNU compiler and OpenMPI 1.6.5. The top row shows the
speedup, i.e. the ratio of the single process total run time and the parallel
run time. The bottom row shows the fraction of the total run time spent in MPI
functions: send and receive operations and idle time due to load imbalances.
We see that the speedup is rather poor, except for small process numbers, and
that the MPI fraction increases significantly with increasing number of
processes. This is to be expected, since load balancing is in no way optimized
in the current version of the code. For a homogeneous set up like the spherical
overdensity this leads to reasonable scaling, but for strongly inhomogeneous
tests like the Evrard collapse this is disastrous.

\subsection{Weak scaling}

Weak scaling is the scaling behaviour of the code when increasing the number of
processes for a fixed problem size per process, i.e. if we double the number of
processes, we double the number of cells in the simulation as well. It is a good
measure for how well the problem is split up over the different processes, as we
expect the amount of work and the amount of communication per process to stay
roughly constant.

\begin{figure}
\centering{}
\includegraphics[width=0.5\textwidth]{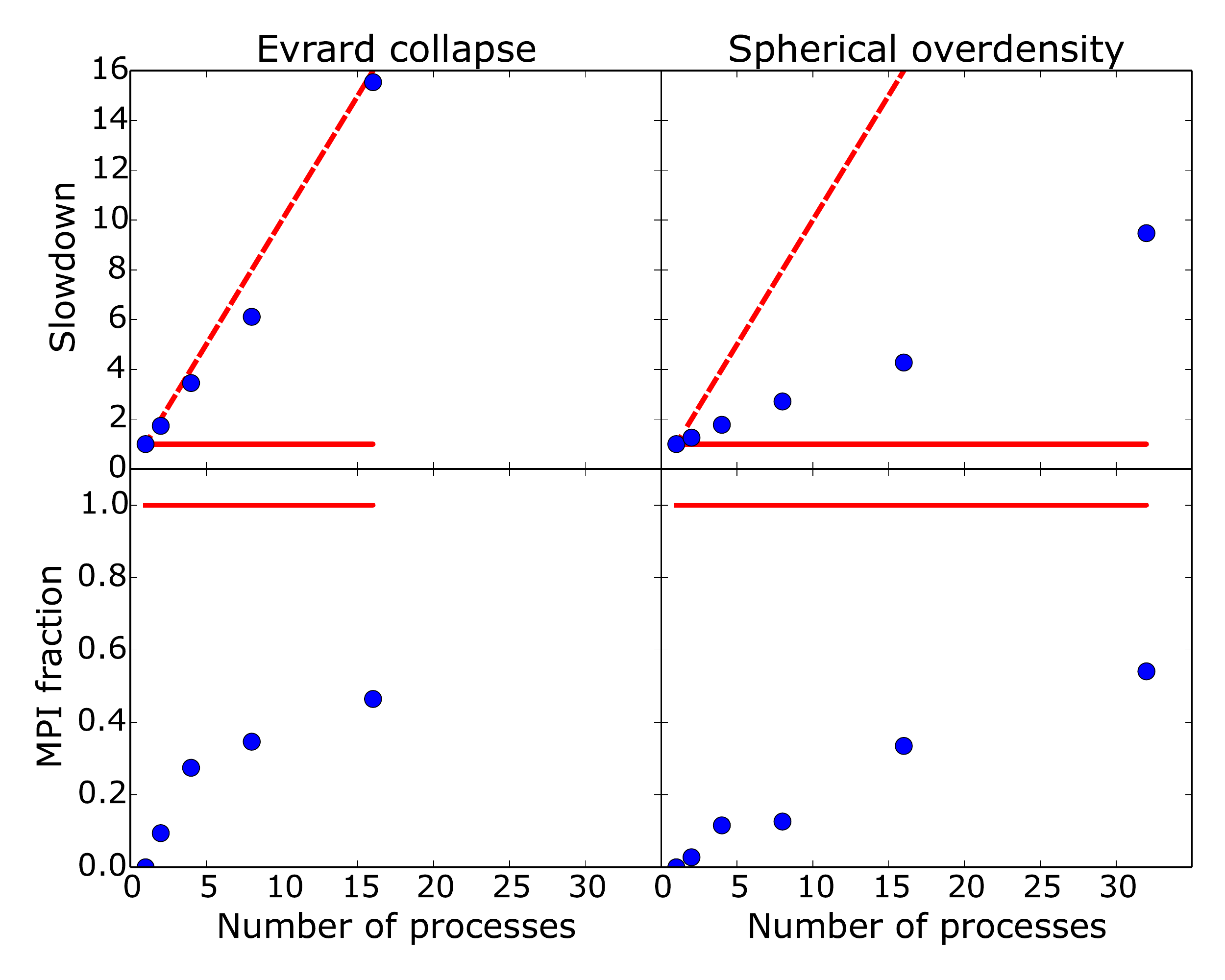}
\caption{Slowdown and MPI fraction for two different tests from the
testsuite under weak scaling. The red dashed line corresponds to a 1:1 relation,
while the full red lines correspond to unity.\label{fig_scaling_weak}}
\end{figure}

\figureref{fig_scaling_weak} shows the weak scaling for the 3D Evrard collapse
and 3D spherical overdensity, with a nominal load of 10,000 cells per process
for both tests. The tests were carried out on the same hardware as the strong
scaling tests discussed above. Instead of the speedup, we now show the slowdown
in the top row, i.e. the ratio of the total parallel run time and the serial run
time. Again, we have rather poor scaling.

\subsection{Components}

An important part of a \textsc{Shadowfax} simulation is the construction of the
Voronoi mesh that is used for the hydrodynamical integration. On average, the
serial version of our code can handle $\approx{}90,000$ Voronoi cells per second
in 2D, and $\approx{}15,000$ Voronoi cells per second in 3D, including the
treatment of reflective or periodic boundaries and the calculation of cell
centroids and volumes, and face midpoints and surface areas.

\begin{figure}
\centering{}
\includegraphics[width=0.5\textwidth]{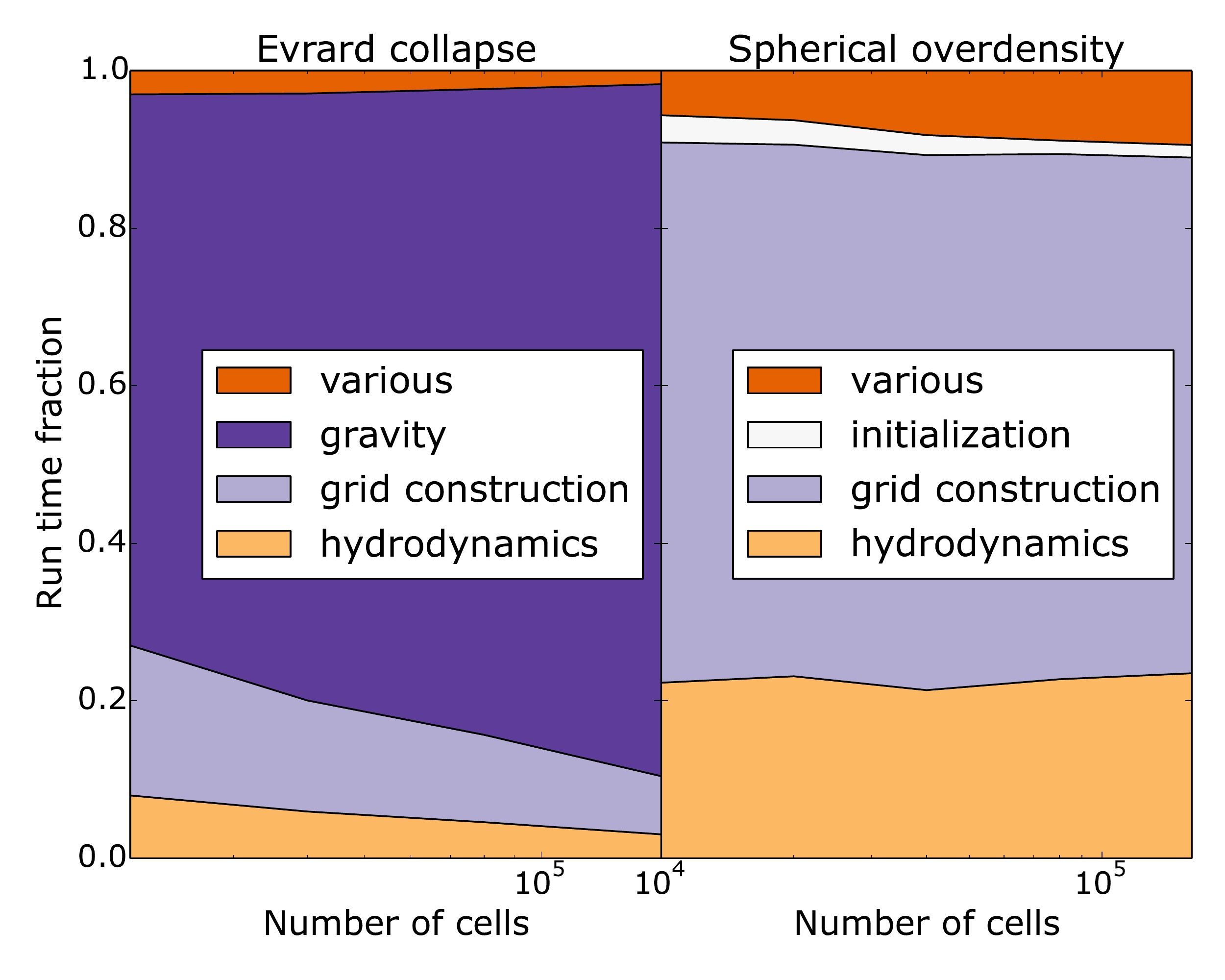}
\caption{Fraction of total run time spent in different parts of the
code.\label{fig_time_components}}
\end{figure}

\figureref{fig_time_components} shows the amount of time spent in various parts
of the code during serial runs with different problem sizes. The current design
of the code makes it very hard to extract the time spent in MPI communications
from the times for the different components, so that we do not show parallel
results.

The gravitational force calculation clearly makes up a large fraction of the
total run time, which increases if the number of cells is increased. Grid
construction in 3D takes up about double the time that is spent in the actual
hydrodynamical integration. Only very little time is spent in other parts of the
algorithm, e.g. particle sorting, tree construction, snapshot output... Program
performance hence does not suffer due to the somewhat less efficient but much
more convenient object oriented design of these parts.

\subsection{Memory consumption}

To test the memory consumption of the code, we used
\textsc{Massif}\footnote{\url{http://valgrind.org/docs/manual/ms-manual.html}},
a heap profiler which is part of the \textsc{Valgrind} instrumentation
framework\footnote{\url{http://valgrind.org/}}. Since we do not expect the
memory imprint to vary much over time, we limited the Evrard test to $t=0.06$
for this test.

\begin{figure}
\includegraphics[width=0.5\textwidth]{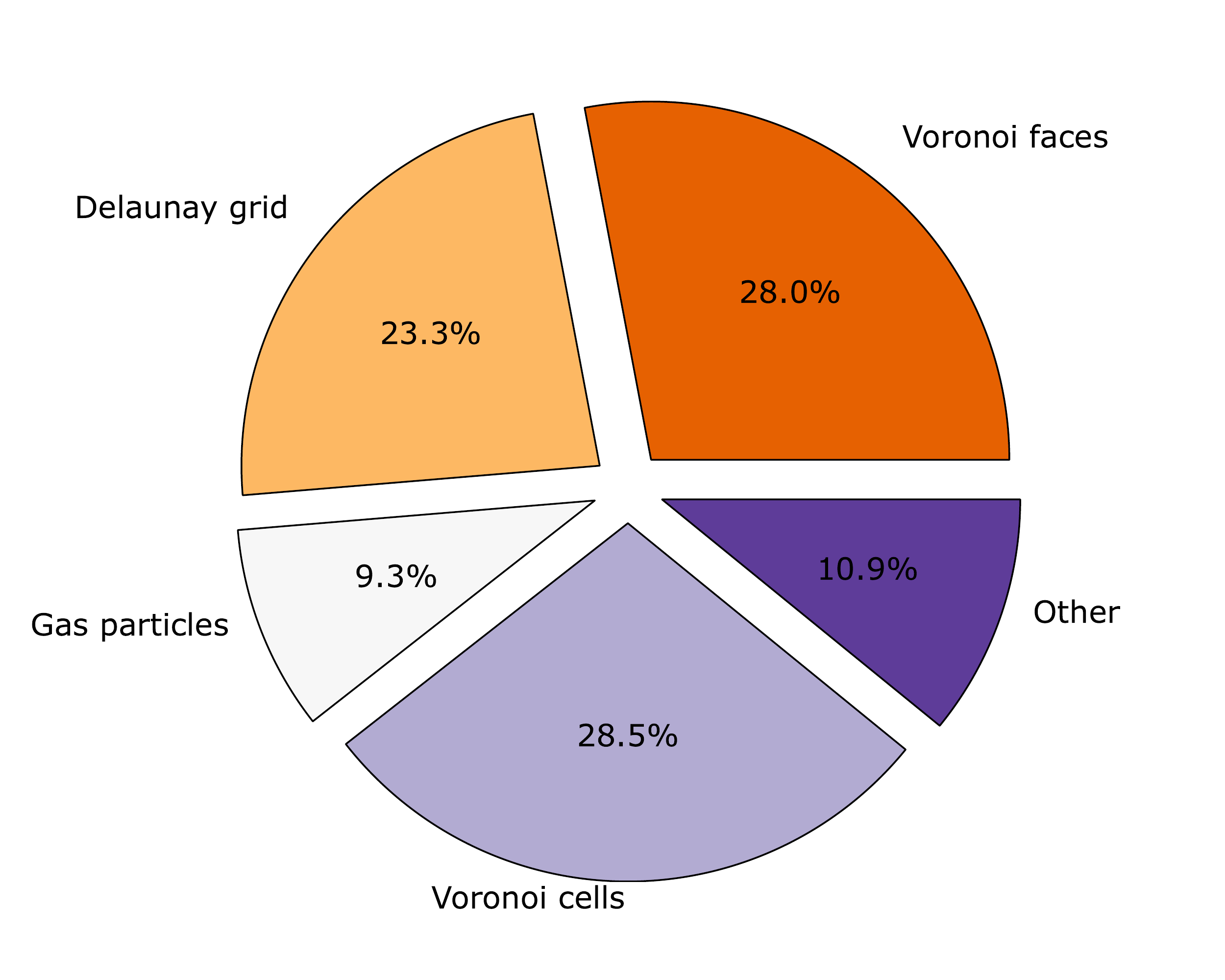}
\caption{The fraction of the memory occupied by various parts of the code during
an Evrard collapse test with 20,000 cells.\label{fig_memory}}
\end{figure}

\figureref{fig_memory} shows the fraction of the memory occupied by various
parts of the code during the final \textsc{Massif} snapshot, when the code used
137 MB of memory in total (including empty blocks used for memory alignment).
A large fraction of this memory is occupied by the Voronoi mesh, with only a
small fraction of the memory used to store the particle properties and the tree
structure needed for the gravity calculation. Note that both the Delaunay
tesselation and the Voronoi mesh are stored simultaneously in the current
version, while the Delaunay tesselation can actually be discarded after the
Voronoi mesh has been constructed. This is left as a future optimization.

\section{Comparison with other methods}

In this section, we compare \textsc{Shadowfax} with a number of other publicly
available hydrodynamical solvers, to qualify the advantages and disadvantages of
the moving mesh method. All files necessary to run these tests are available
from \url{http://www.dwarfs.ugent.be/shadowfax/}.

\subsection{Kelvin-Helmholtz instabilities}

It is a known problem of particle-based Lagrangian methods like SPH that they
have difficulties resolving Kelvin-Helmholtz instabilities \citep{2007Agertz},
since the basic SPH equations do not cover discontinuous solutions, like shock
waves and contact discontinuities. As \citet{2010Springel} showed, a moving mesh
method does not experience these difficulties, as the Riemann problem based
finite volume method that is used does include discontinuous solutions.

One of the main advantages of a Lagrangian method over methods that use a fixed
discretization (like AMR), is the Galilean invariance of the method. This means
that the flux between two neighbouring cells will only depend on the relative
velocity of two neighbouring cells, and not on their absolute velocities with
respect to some reference frame fixed to the simulation box. As a result, we
expect a moving mesh method to better resolve instabilities in a fluid that is
moving with a high bulk velocity with respect to the simulation box reference
frame. To quantify this behaviour, we set up a variant of the shearing layers
test introduced in \citet{2007Agertz}.

For this test, two layers which can have different densities are in pressure
equilibrium inside a periodic box. The layers receive a velocity component
parallel to the interface between the layers, but with opposite sign, so that
they shear against each other. Due to the pressure equilibrium, the system is
marginally stable~: a small velocity component perpendicular to the interface
between the layers will exponentially grow to form a Kelvin-Helmholtz
instability. The instability first goes through a linear phase of exponential
growth, after which the system becomes highly non-linear. The simple setup
described here is unstable on all scales, so that instabilities can even be
seeded by numerical noise in the absence of diffusion. As a result, the results
are highly dependent of the resolution and the details of the numerical scheme.
This makes a thorough comparison of different methods completely impossible
\citep{2016Lecoanet}.

As \citet{2014Hendrix} point out, introducing a middle layer with a linear
transition in flow velocity in between the shearing layers will suppress small
scale instabilities, so that the growth of instabilities no longer depends on
the numerical resolution (if it is high enough). The middle layer also
introduces a maximally unstable wavelength, which we will seed. This gives us
full control over the instabilities that will grow. We first discuss the
non-linear growth of the instability in a setup with a density contrast of $10$
between the layers, to show that \textsc{Shadowfax} qualitatively produces
similar instabilities. We then study the convergence of the linear growth rate
in a setup without density contrast.

\subsubsection{Bulk velocity}

We set up a periodic 2D box with unit length, in which the density is given by
\begin{equation}
\rho{}(x, y) = \begin{cases}
1 & y < 0.25 \\
10 & 0.25 \leq{} y \leq{} 0.75 \\
1 & 0.75 < y.
\end{cases}
\end{equation}
The $x$ component of the velocity is given by
\begin{equation}
v_x = \begin{cases}
-0.5 & y \leq{} 0.25 - d \\
- 0.5 + \frac{y + d - 0.25}{2d} & 0.25-d < y < 0.25 + d \\
\phantom{-}0.5 & 0.25 + d \leq{} y \leq{} 0.75 - d \\
\phantom{-}0.5 -\frac{y + d - 0.75}{2d} & 0.75 - d < y < 0.75 + d \\
-0.5 & 0.75 + d \leq{} y,
\end{cases}
\label{eq_vx}
\end{equation}
with $d=0.025$ the thickness of the middle layer. The $y$ component of the
velocity is
\begin{equation}
v_y = A \sin{}(4\pi{}x) \left( \text{e}^{-\frac{(y-0.25)^2}{2\sigma{}^2}} +
\text{e}^{-\frac{(y-0.75)^2}{2\sigma{}^2}} \right),
\end{equation}
with $A=0.1$ and $\sigma{} = 0.00125$. The pressure in the entire box is set to
the constant value $p=2.5$.

We compare the results obtained with \textsc{Shadowfax}, with results obtained
using the AMR code MPI-AMRVAC\footnote{ascl:1208.014} \citep{2012Keppens}, using
the same initial condition (we use a Cartesian initial grid for the
\textsc{Shadowfax} runs). MPI-AMRVAC supports different hydrodynamical schemes;
we use both a conservative finite difference scheme with global Lax-Friedrich
splitting and a fifth order spatial reconstruction (hereafter called FD), and a
finite volume scheme with a Harten-Lax-van Leer Contact (HLLC) solver (hereafter
called FV). Both schemes use a fourth order accurate Runge-Kutta time
integration scheme.

\begin{figure*}
\centering{}
\includegraphics[width=\textwidth]{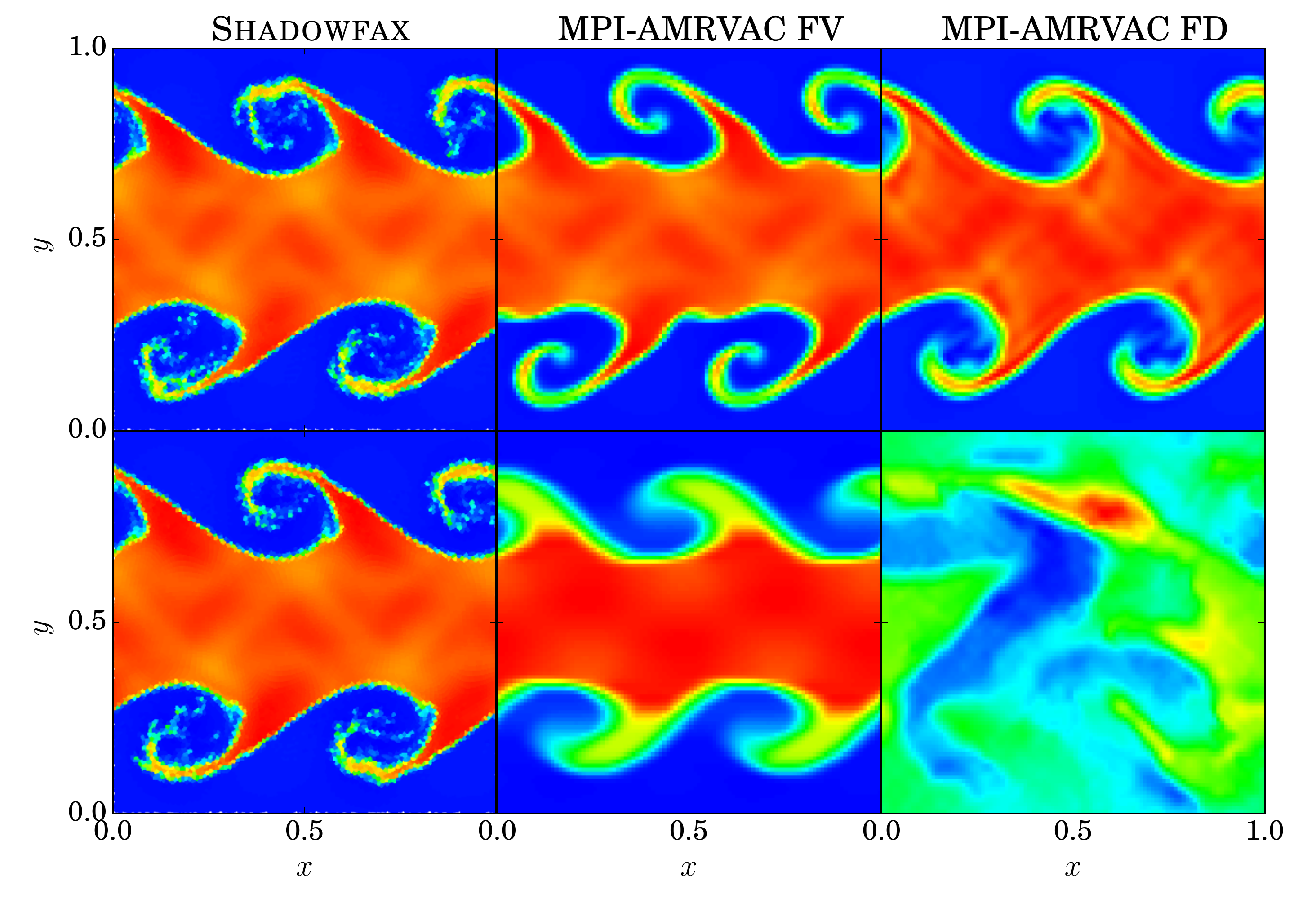}
\caption{Density colour plot for the shearing layers test at time $t=1.5$. The
top row corresponds to simulations with $v_\text{bulk}=0$, while the bottom row
corresponds to simulations with $v_\text{bulk}=100$. All simulations start from
a $100\times{}100$ Cartesian grid and have a fixed number of cells. The
individual cells are shown, this explains the irregularities at the boundaries
of the \textsc{Shadowfax} plot.\label{fig_kh_lowres}}
\end{figure*}

\begin{figure}
\centering{}
\includegraphics[width=0.5\textwidth]{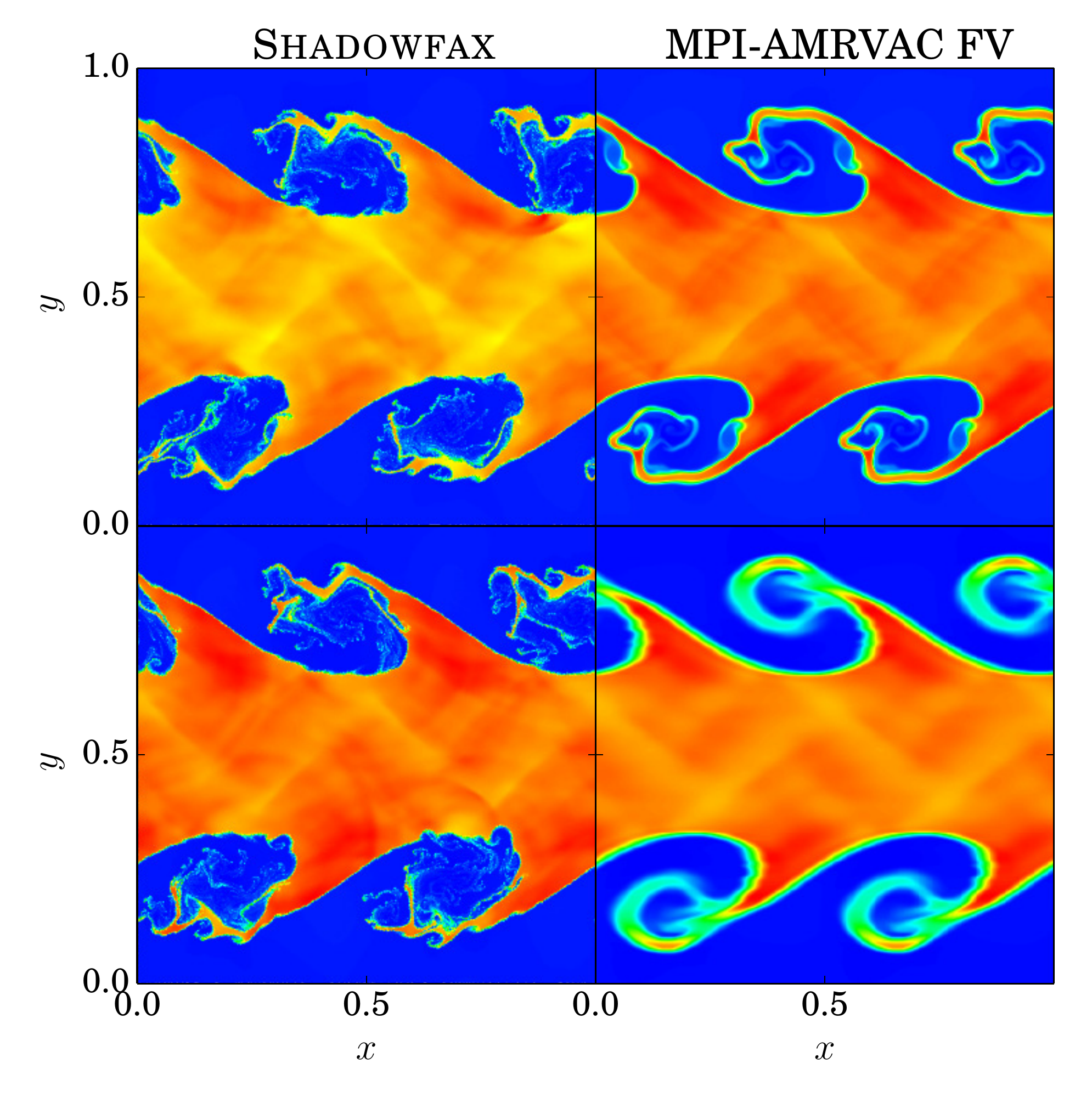}
\caption{Density colour plot for the shearing layers test at time $t=1.5$ for
simulations using a $400\times{}400$ grid. The top row corresponds to
simulations with no bulk velocity, the bottom row has
$v_\text{bulk}=100$.\label{fig_kh_hires}}
\end{figure}

To test the Galilean invariance of the code, we optionally add a bulk velocity
$v_\text{bulk}=100$ to the entire fluid (corresponding to a Mach number of 155
in the high density layer), so that the fluid moves with respect to the
simulation box reference frame. The results of a low resolution run at time
$t=1.5$ are shown in \figureref{fig_kh_lowres}. Without a bulk velocity, the
three methods produce similar results, in the sense that they all produce the
same large instabilities on the same timescale. There are some differences in
the non-linear phase of the instability, which is to be expected. With a bulk
velocity, the results dramatically change for the MPI-AMRVAC simulations. The FD
method does not produce any result at all, while the instabilities in the FV
simulation are clearly affected by the bulk velocity. There is also a clear
imprint on the run time of the MPI-AMRVAC simulations, since a much smaller
system time step is needed for the integration. The run time and the results of
the \textsc{Shadowfax} simulations are not affected by the bulk velocity, since
the mesh is moving along with the flow. These results are confirmed by the high
resolution runs, shown in \figureref{fig_kh_hires}.

\subsubsection{Linear growth rate}

In a setup without density contrast, the initial exponential growth of the
instability only depends on the wavelength of the instability and the thickness
of the middle layer \citep{2014Hendrix}. This means that simulations of this
initial phase should converge to the same growth rate, irrespective of the
resolution or the method that is used.

To test this, we run a variant of the shearing layers test without density
contrast. We still use a periodic box with unit length, but now the density and
pressure in the box are both constant and equal to $1$. The $x$ component of the
velocity is given by \eqref{eq_vx}, while the $y$ component of the velocity is
now given by
\begin{equation}
v_y = B \sin{}(4\pi{}x) \left( \text{e}^{-\frac{(y-0.25)^2}{32d^2}} +
\text{e}^{-\frac{(y-0.75)^2}{32d^2}} \right),
\end{equation}
with $B=0.0005$ and $d=0.0317$.

To quantify the growth of the instability, we track the total kinetic energy in
the $y$ direction. Initially, this energy is set by our seed velocity, but as
the instability grows, kinetic energy in the $x$ direction is converted into
extra kinetic energy in the $y$ direction, so that this energy will grow
exponentially, as does the $y$ component of the velocity.

\begin{figure}
\centering{}
\includegraphics[width=0.5\textwidth]{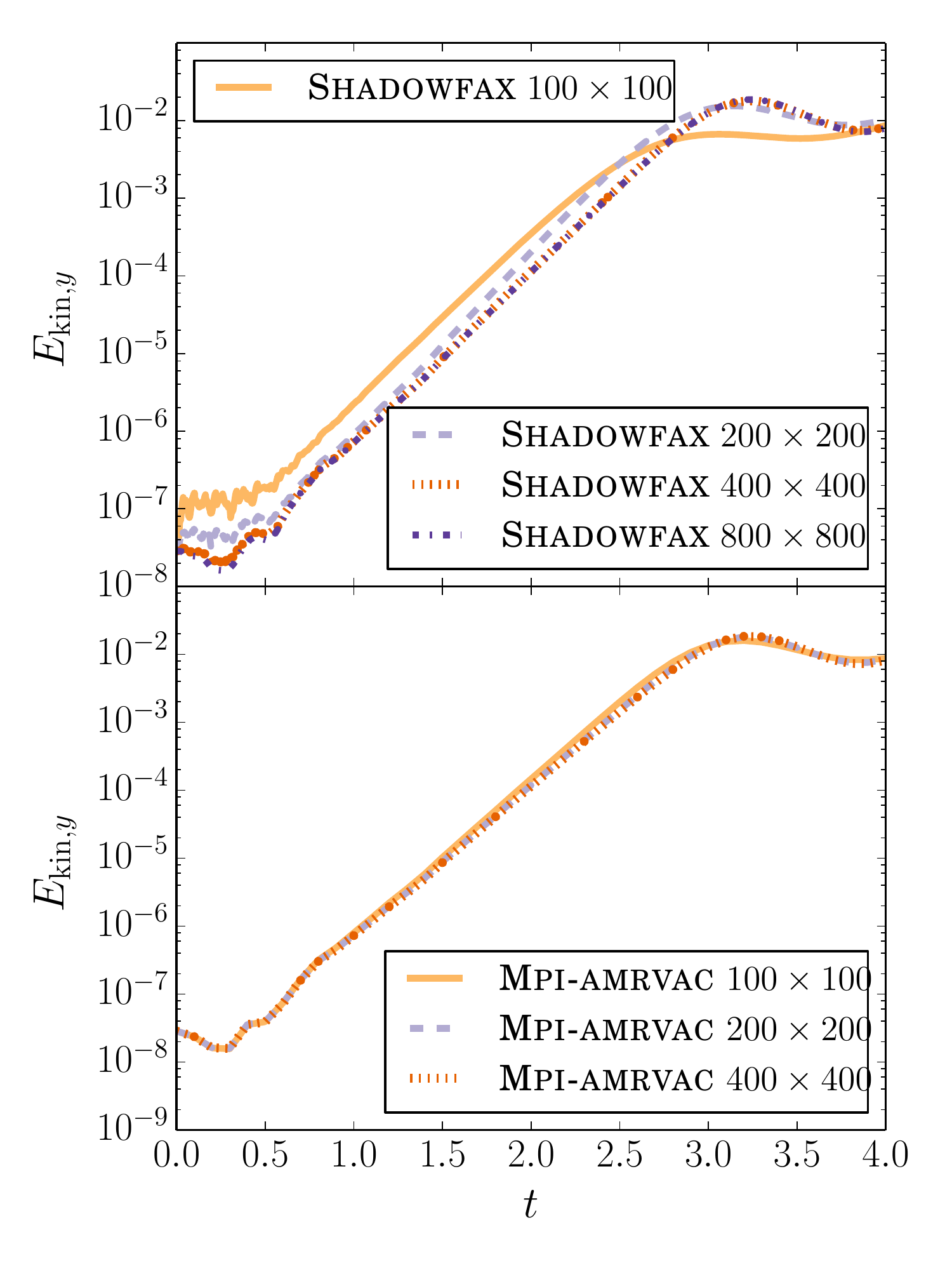}
\caption{The kinetic energy in the $y$ direction as a function of time for the
shearing layers test without a density contrast.\label{fig_growth_rate}}
\end{figure}

\begin{figure}
\includegraphics[width=0.5\textwidth]{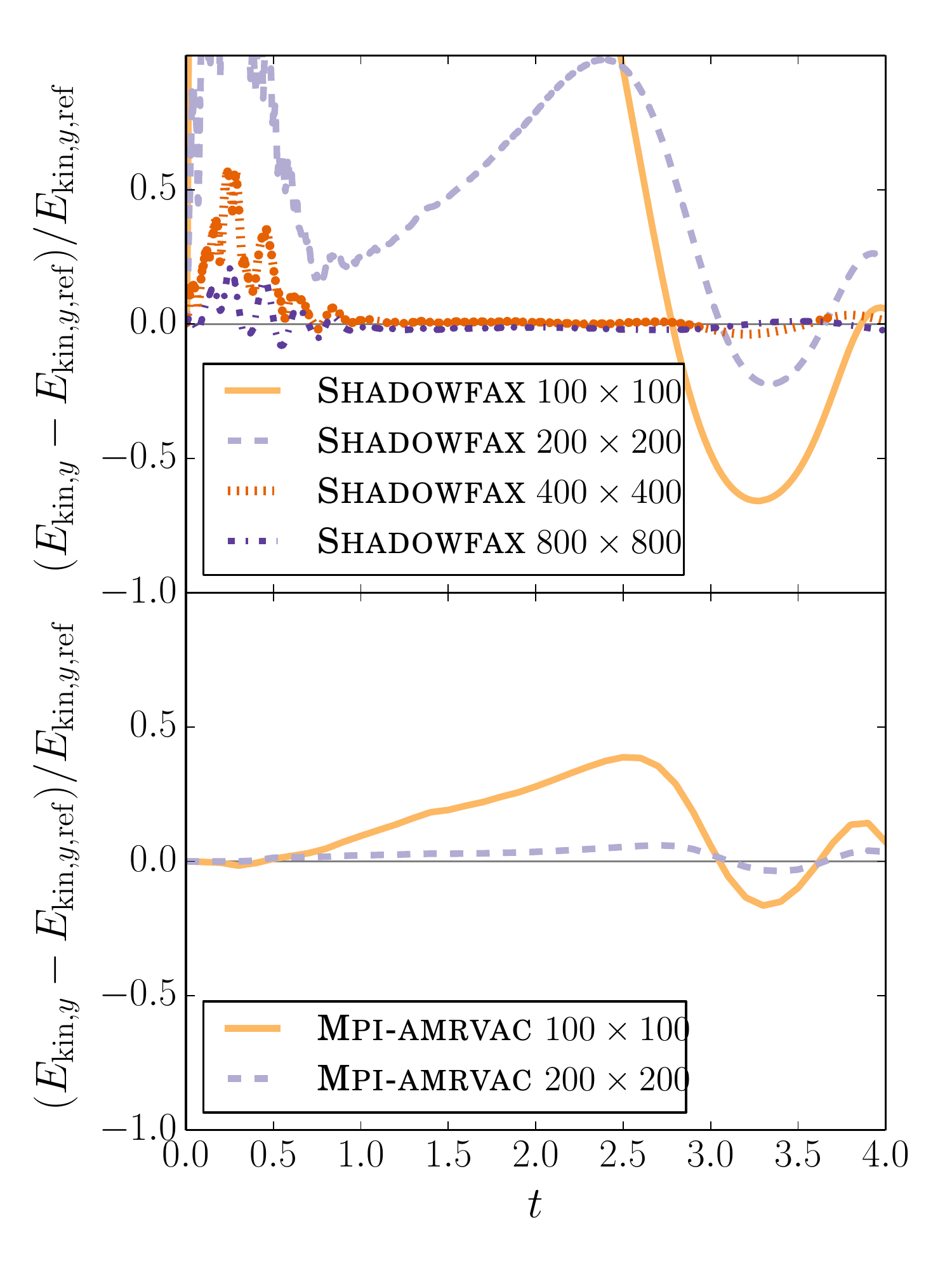}
\caption{The relative difference between the kinetic energy in the $y$ direction
for the $400\times{}400$ MPI-AMRVAC simulation and that for the other
simulations, as a function of time.\label{fig_growth_rate_convergence}}
\end{figure}

\figureref{fig_growth_rate} shows the kinetic energy in the $y$ direction as a
function of time for our simulations and for grids with different resolutions.
The high resolution \textsc{Shadowfax} results are in good agreement with the
high resolution MPI-AMRVAC results, but the convergence is slower for
\textsc{Shadowfax}. This is illustrated in
\figureref{fig_growth_rate_convergence}, where we plot the relative difference
between the different simulations and the high resolution MPI-AMRVAC result,
that we use as a reference solution.

\begin{table}
\centering{}
\caption{Slope of an exponential fit to the kinetic energy in the $y$ direction
in the time interval $[1.5,2.5]$ for the shearing layers tests without density
contrast.\label{tab_growth_rate}}
\begin{tabular}{r c}
\hline
simulation & slope \\
\hline
\textsc{Shadowfax} $100\times{}100$ & 4.61 \\
\textsc{Shadowfax} $200\times{}200$ & 5.43 \\
\textsc{Shadowfax} $400\times{}400$ & 5.10 \\
\textsc{Shadowfax} $800\times{}800$ & 5.10 \\
\hline
MPI-AMRVAC $100\times{}100$ & 5.28 \\
MPI-AMRVAC $200\times{}200$ & 5.14 \\
MPI-AMRVAC $400\times{}400$ & 5.12 \\
\hline
\end{tabular}
\end{table}

We fitted an exponential function of the form $B \text{e}^{At}$ to the kinetic
energy in the $y$ direction, in the time interval $[1.5, 2.5]$, and use the
slope $A$ to quantify the growth of the instability. The results are shown in
\tableref{tab_growth_rate}. Both the high resolution \textsc{Shadowfax} and
MPI-AMRVAC results are converged.

\subsection{\textsc{Shadowfax} versus \textsc{swift}}

The mesh-free methods introduced by \citet{2015Hopkins} use the same finite
volume method used by \textsc{Shadowfax}, but use an SPH-like discretization of
the fluid to calculate volumes and interfaces between neighbouring particles.
Since this method does not require the construction of a global unstructured
mesh, it is computationally cheaper and has a better potential for parallel
scalability. Just as SPH however, the method smooths out local resolution over a
number of neighbouring particles, potentially lowering the effective resolution.

We compare \textsc{Shadowfax} with our own implementation of a mesh-free method
in the highly parallel SPH-code
\textsc{swift}\footnote{\url{
https://gitlab.cosma.dur.ac.uk/swift/swiftsim/tree/gizmo}}.
This method combines the fast neighbour loop algorithms of \textsc{swift} with a
finite volume method that is the same as implemented in \textsc{Shadowfax}, and
uses the same exact Riemann solver. Since \textsc{Shadowfax} reads the same
initial condition file format as \textsc{swift}, we can use the exact same
initial condition for both simulations and directly compare the results. We also
compare with the default SPH version of \textsc{swift}.

\begin{figure*}
\centering{}
\includegraphics[width=\textwidth]{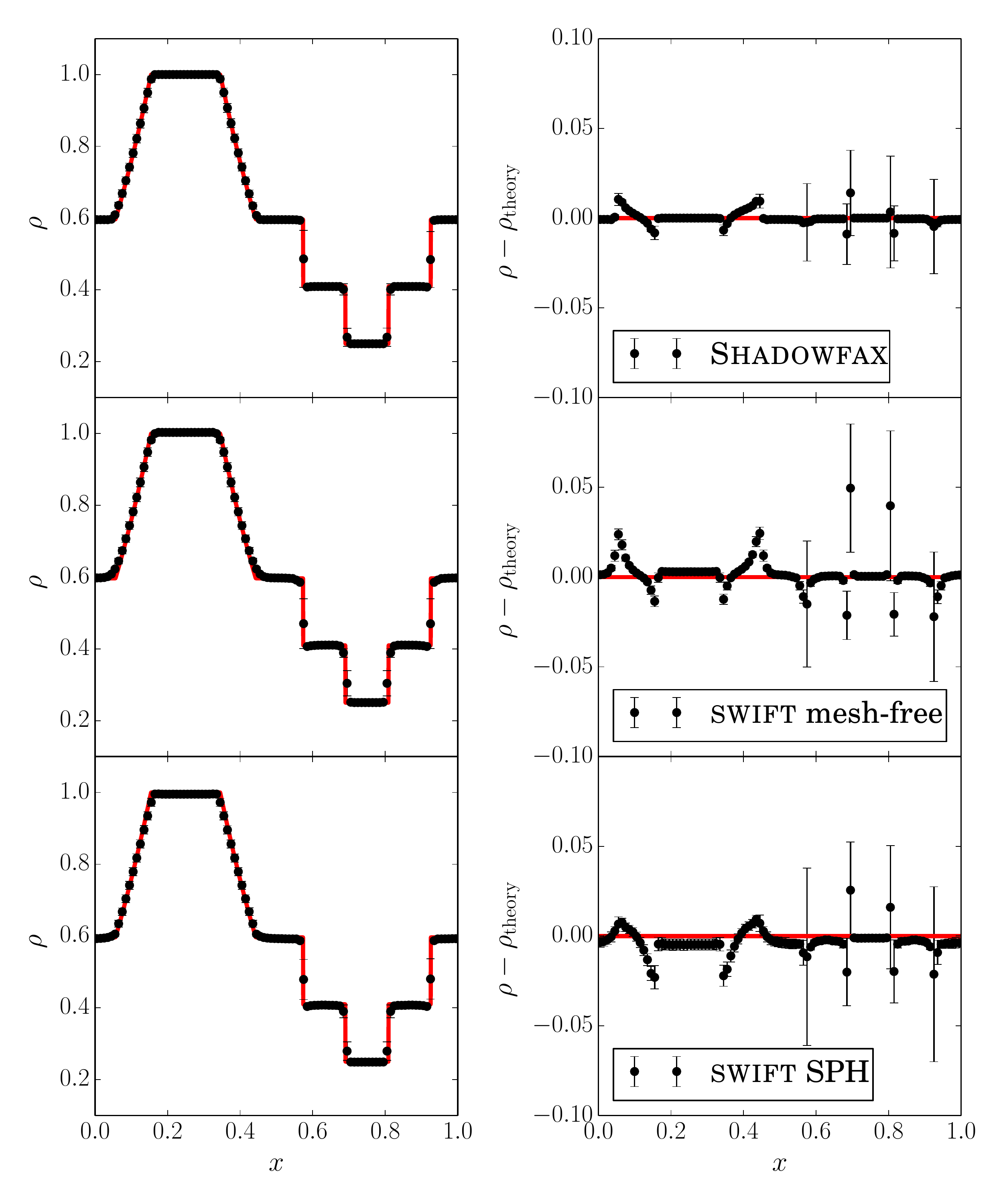}
\caption{Density profile of the Sod shock test at $t=0.12$. The black dots
represent the binned density values, with the error flags indicating the
standard deviation on the values within the bins. The red line corresponds to
the exact solution, which is the solution of the equivalent Riemann
problem. The right column shows the difference between the density and the
analytical solution. All simulations use the same initial condition with
1,024,128 particles.\label{fig_gizmo_sod}}
\end{figure*}

\figureref{fig_gizmo_sod} shows the density profile of a Sod shock test, which
is the 1D equivalent of the spherical overdensity test discussed as part of the
\code{testsuite}. A left high density, high pressure region with $\rho{}=1$
and $p=1$ connects with a low density, low pressure region with $\rho{}=0.25$
and $p=0.1795$ at $x=0.5$ inside a cuboid with dimensions
$1\times{}0.125\times{}0.125$ with periodic boundaries. The results are evolved
to time $t=0.12$, when the left rarefaction wave, central discontinuity and
right shock have developed.

The \textsc{Shadowfax} result clearly follows the theoretical curve, while both
\textsc{swift} results show minor deviations around the different features in
the profile. The deviations are strongest for the \textsc{swift} mesh-free
results, although the noise on the \textsc{swift} SPH result is higher. We
calculated $\chi{}^2$ values for all three simulations by summing the quadratic
differences between the particle densities and the theoretical densities for all
particles. This yielded $\chi{}^2=29.96$ for \textsc{Shadowfax},
$\chi{}^2=106.71$ for \textsc{swift} mesh-free, and $\chi{}^2=123.54$ for
\textsc{swift} SPH. Not smoothing out the local resolution hence clearly leads
to an overall higher accuracy of the moving mesh method.

\subsection{Noh test}

The strong shock test proposed by \citet{1987Noh} is a very challenging test
with a known analytical solution. It consists of a reflective box with unit
length, in which a fluid with unit density and a negligible thermal energy of
$1\times{}10^{-5}$ is enclosed. The radial velocity of the fluid is set to $-1$
at $t=0$, so that the fluid collapses on the origin and causes a strong shock
with a very high Mach number.

The velocity of the shock front is $v_\text{shock}=1/3$, and the radial
density profile at time $t>0$ is given by
\begin{equation}
\rho{}(r, t) = \begin{cases}
16 & r \leq{} v_\text{shock}t \\
1 + \frac{t}{r} & v_\text{shock} t < r
\end{cases}
\end{equation}
in 2D.

As is common for this problem, we restrict ourselves to the upper right quadrant
of the box. However, we do not use the commonly used inflow boundary conditions
for the upper and right boundaries \citep{2010Springel}, since
\textsc{Shadowfax} does not currently support inflow boundaries. Since the
radial flow of the fluid creates a very low density cavity at these boundaries,
this leads to numerical problems when the shock reaches the low density
cavities. We will therefore restrict the simulation to $t=0.5$ and use a higher
resolution compared to \citet{2010Springel}, so that the lower left quarter of
our box can be compared with his result. We only show the 2D version of the
test. For comparison, we also run the same test using MPI-AMRVAC.

\begin{figure*}
\centering{}
\includegraphics[width=\textwidth]{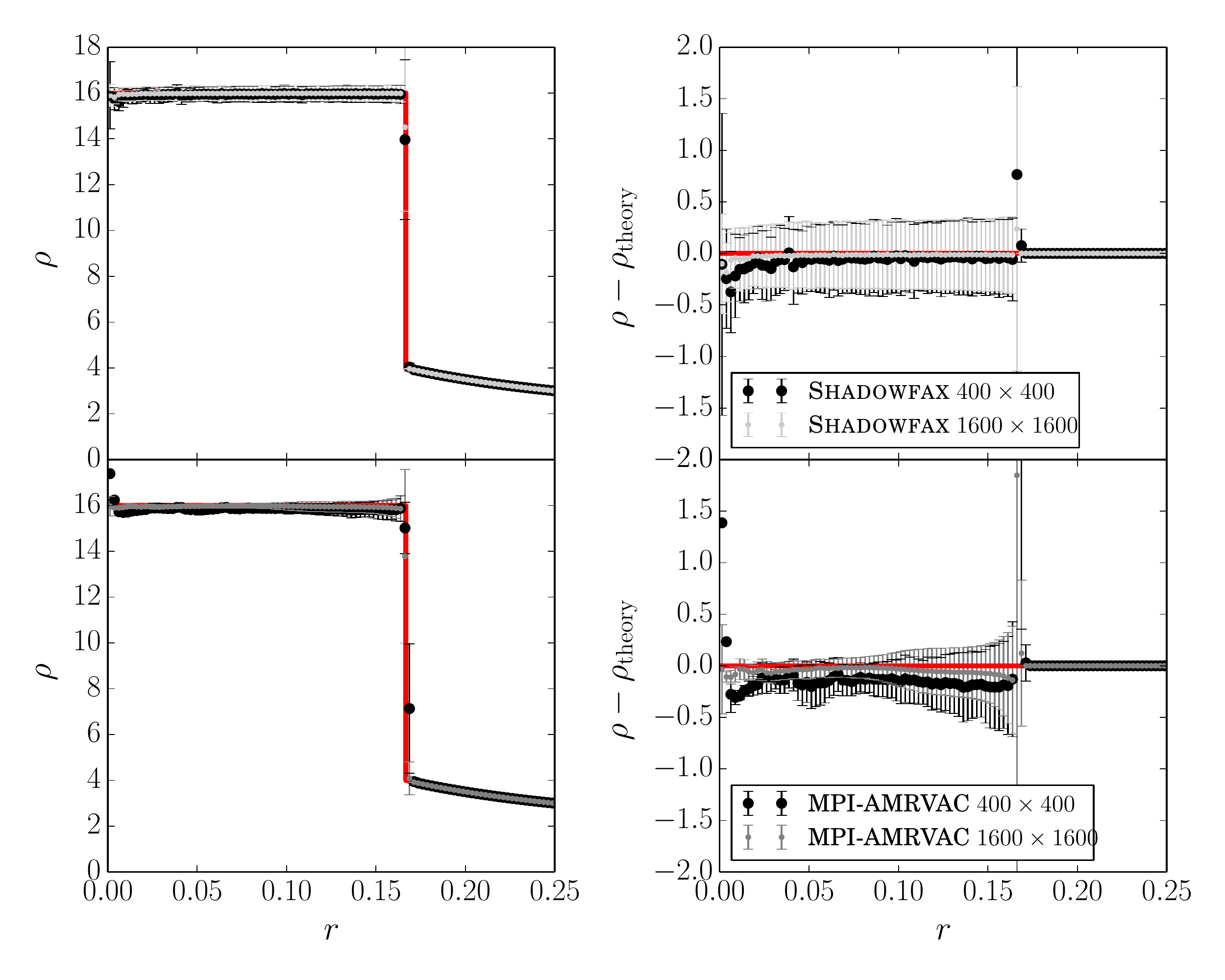}
\caption{Radial density profile for the Noh test at $t=0.5$. The black and gray
dots show the binned density values, with the error bars indicating the standard
deviation of the values inside the bins. The red line corresponds to the
analytical solution. The right column shows the difference between the density
and the analytical solution.\label{fig_noh}}
\end{figure*}

\figureref{fig_noh} shows the radial density profile at $t=0.5$, together with
the analytical solution. The results for both methods are in good agreement, but
the low resolution MPI-AMRVAC result has a density peak at the origin which is
less pronounced in the \textsc{Shadowfax} result. We calculated $\chi{}^2$
values for the simulations, yielding $\chi{}^2=1.78\times{}10^4$ and
$\chi{}^2=1.55\times{}10^5$ for \textsc{Shadowfax}, and
$\chi{}^2=2.87\times{}10^5$ and $\chi{}^2=4.54\times{}10^6$ for MPI-AMRVAC.

\subsection{Implosion test}

Another challenging test is the implosion test of \citet{2003Liska}. It consist
of a periodic box of length $0.6\times{}0.6$, in which a fluid with unit density
and pressure is initially at rest. In the center of the box, a rhombus with
width $0.3$ is cut out, in which the density and pressure are given lower
values~: $\rho{}=0.125$, $p=0.14$. The high density fluid will implode into this
region and cause a strong shock wave, that will travel back and forth in the
periodic box. As in \citet{2003Liska}, we set the adiabatic index $\gamma{}=1.4$
for this test.

This test involves both a large scale strong shock wave, and a lot of small
scale instabilities at the interface between high and low density region, which
interact with this shock wave. Since the initial conditions do not contain
middle layers that could suppress small scale instabilities, the results in
general will not converge. The behaviour of the strong shock wave is however
similar for different methods, as \citet{2003Liska} showed. It is furthermore
interesting to see whether a code can handle this complex problem.

\begin{figure*}
\centering{}
\includegraphics[width=0.8\textwidth]{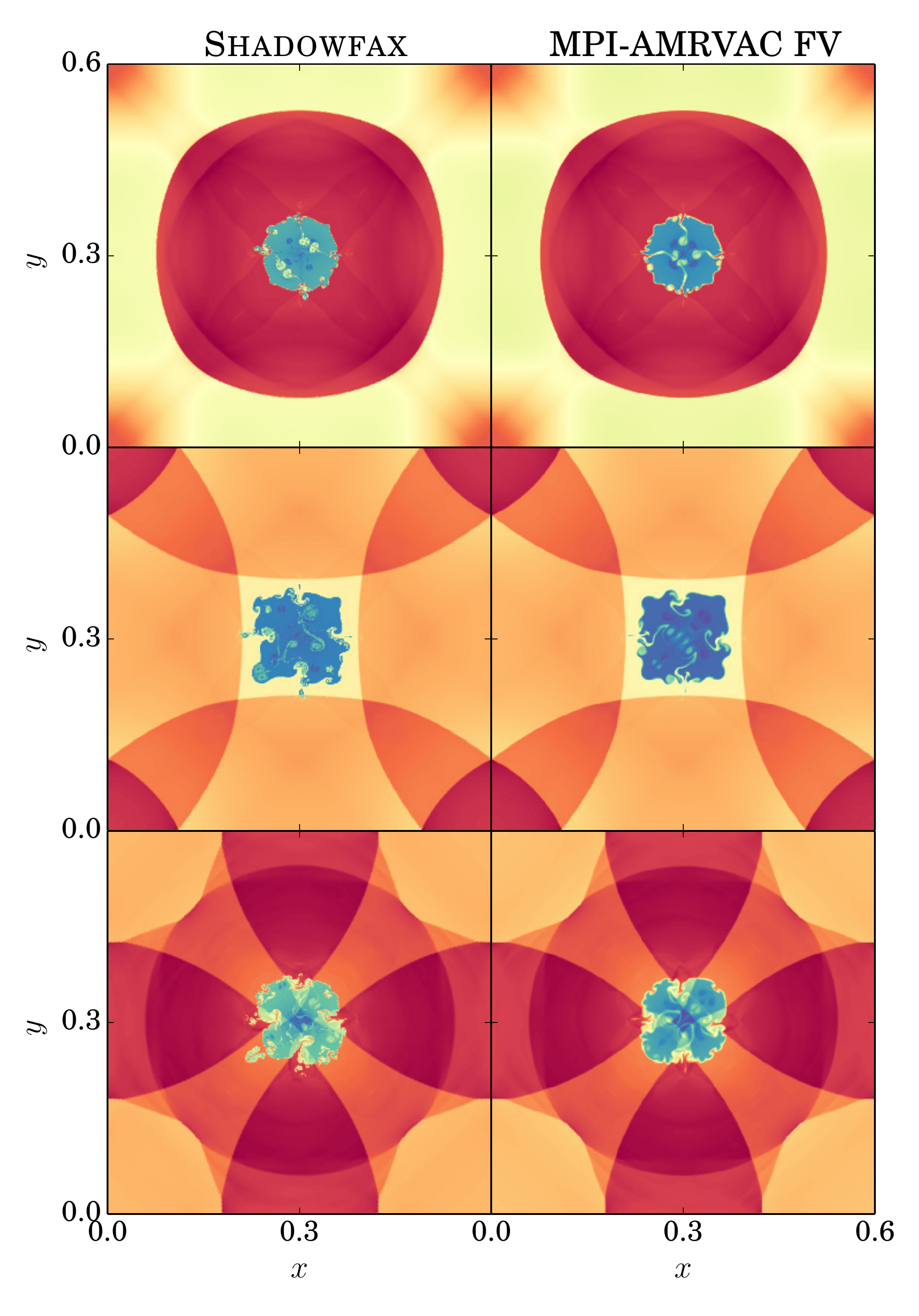}
\caption{Density colour plots for the implosion test at different times.
\emph{Top:} $t=0.25$, \emph{middle:} $t=0.5$, \emph{bottom:}
$t=0.75$.\label{fig_implosion}}
\end{figure*}

In \figureref{fig_implosion}, we show some snapshots of the implosion test and
compare \textsc{Shadowfax} results with results obtained using MPI-AMRVAC. It is
interesting to see that both methods produce asymmetrical results in the center,
while reproducing the same large scale shock wave, even at later times, when the
shock wave has interacted with the central instabilities. Just as before, the
\textsc{Shadowfax} instabilities seem to develop faster than the ones in the
MPI-AMRVAC result, and they are also significantly more chaotic, due to the
co-moving character of the mesh. For the same mesh resolution, the
\textsc{Shadowfax} results show more instabilities, indicating that a moving
mesh has a higher local resolution. \figureref{fig_implosion_zoom} shows a zoom
of the \textsc{Shadowfax} result at time $t=0.5$ with the mesh overplotted, to
illustrate how the mesh adapts to the local density.

We must however stress that the setup of this problem makes it impossible to
compare these instabilities, and that it is impossible to determine if these
instabilities are converged in any way. It is hence impossible to make a claim
about which method is better. We can only point out the differences between both
methods.

\section{Discussion}

In this paper, we introduced the public simulation code \textsc{Shadowfax}. The
code can be used to simulate a mixture of gas and cold dark matter, with an
accurate treatment of the hydrodynamics and gravitational forces. The
discretization of the gas is provided by an unstructured Voronoi mesh, which is
evolved in time to more accurately follow the hydrodynamical flow.

The code is inspired by \citet{2010Springel}, but uses an object-oriented
design, which attempts to make the code easier to read and extend. Some parts of
the code make use of advanced C++ language features, like templates, to improve
code reuse without significant run time cost. We have attempted to separate the
actual physics from the algorithmic details as much as possible. Much
improvement is possible however, which will be the subject of future work on the
code.

We have introduced the test problems that are currently included in the public
version of the code, as well as some more involved tests that compare the code
with other publicly available codes. These tests show that the Lagrangian nature
of the moving mesh method makes it better at resolving instabilities in regions
with high Mach numbers than fixed grid methods. On the other hand, a moving mesh
is a lot more sensitive to instabilities that are seeded numerically. When the
growth rate of instabilities is studied, a moving mesh method requires higher
resolution to converge to a consistent growth rate than a fixed grid method.

The finite volume method used by \textsc{Shadowfax} is almost identical to that
used by the mesh-free methods of \citet{2015Hopkins}, but has a higher effective
resolution for the same number of particles, since the mesh-free discretization
of space smooths out resolution over a large number of neighbours.

\section*{Acknowledgments}

We thank the two anonymous referees for the constructive feedback that improved
the quality of this manuscript, and for pointing out the details of our
implementation we forgot to mention in the first version.
We thank the Ghent University Special Research Fund and the Interuniversity
Attraction Poles Programme initiated by the Belgian Science Policy Office (IAP
P7/08 CHARM) for financial support. We thank Robbert Verbeke for his
contribution to the exact Riemann solver, and Peter Camps for sharing the
\code{Vec} class that originally was part of SKIRT\footnote{ascl:1109.003} with
us, before SKIRT became public. BV thanks Matthieu Schaller and Tom Theuns for
the discusions about finite volume methods, and the coupling between
hydrodynamics and gravity. We thank Volker Springel and Romain Teyssier for
making their codes \textsc{Gadget2} and \textsc{ramses} public. Special thanks
also to Rony Keppens for sharing his knowledge about grid methods and helping us
setting up the MPI-AMRVAC simulations. Finally, we want to thank J. R. R.
Tolkien for providing an extensive list of unique names to give to computing
infrastructure and simulation codes.

\section*{References}
\bibliography{main}{}
\bibliographystyle{elsarticle-harv}

\end{document}